\documentclass[aps,preprint,nofootinbib,floatfix]{revtex4-2}
\usepackage[utf8]{inputenc}
\usepackage{subfigure}
\usepackage[graphicx]{realboxes}
\usepackage{graphicx,tabularx}
\usepackage{float}
\usepackage{setspace}
\usepackage{hyperref}
\usepackage{bm}
\usepackage{multirow}
\usepackage{amsthm,amsmath,amssymb}
\usepackage{mathrsfs}
\usepackage{amsmath}
\usepackage{indentfirst}
\usepackage{tikz}
\usepackage{url}
\usepackage{footmisc}
\usepackage{amsfonts}
\usepackage{cancel}
\usepackage{color}
\usepackage{multirow}
\usepackage{slashed}
\usepackage{comment}
\usepackage{appendix}
\usepackage{soul}
\usepackage{setspace}
\usepackage{titletoc}
\usepackage{booktabs}

\newcommand{\HL}[1]{{\color{blue} [HL: #1]}}
\newcommand{\LZY}[1]{{\color{orange} [LZY: #1]}}
\begin{document}
\title{Phenomenology of Heavy Neutral Gauge Boson at Muon Collider}
\author{Zongyang Lu$^{a}$}
\author{Honglei Li$^{a}$}
\email{sps\_lihl@ujn.edu.cn}
\author{Zhi-Long Han$^{a}$}
\email{sps\_hanzl@ujn.edu.cn}
\author{Zong-Guo Si$^{b}$}
\email{zgsi@sdu.edu.cn}
\author{Liuxin Zhao$^{a}$}
\affiliation{$^a$School of Physics and Technology, University of Jinan, 250022, Jinan, China}
\affiliation{$^b$School of Physics, Shandong University, 250100, Jinan, China}
\date{\today}
\begin{abstract}
Heavy neutral gauge boson $Z^\prime$ is proposed in many new physics models.  It has rich phenomena at the future muon collider. We study the properties of $Z^\prime$ boson   with the process of  $\mu^+ \mu^- \rightarrow q \bar{q}$, $\mu^+ \mu^- \rightarrow l^+ l^-$, $\mu^+ \mu^- \rightarrow Z H$ and $\mu^+ \mu^- \rightarrow W^+ W^-$. The discrepancy of  $Z^\prime$  coupling to different types of  particles can be shown in the cross section distributions around the resonance peak  of various decay modes.
Angular distributions of the final quark or lepton in  $\mu^+ \mu^- \rightarrow q \bar{q}/l^+ l^- $ process are sensitive to the parameters such as mass of $Z^\prime$ and the $Z-Z^\prime$ mixing angle. The interaction of new gauge boson coupling to the standard model gauge particles and Higgs boson are also studied through $\mu^+ \mu^- \rightarrow Z H \rightarrow l^+l^- b \bar{b}$ and $\mu^+ \mu^- \rightarrow W^+W^-  \rightarrow l^+l^- \nu_l  \bar{\nu}_l$. The cross section and the final particles' angular distributions with the contribution of $Z^\prime$ boson differ from those processes with only  standard model particles. A forward-backward asymmetry defined by the angular distribution is provided to show the  potential of searching for new physics at the muon collider.  Especially, the beam polarization with certain value can effectively enlarge the forward-backward asymmetry.\\



\end{abstract}

\maketitle
\newpage
\quad

\titlecontents{section}[0pt]{\addvspace{5pt}\filright}              
{\contentspush{\thecontentslabel\ 
}}              
{}{\titlerule*[6pt]{.}\contentspage}
\titlecontents{subsection}[15pt]{\addvspace{5pt}\filright}              
{\contentspush{\thecontentslabel\ 
}}              
{}{\titlerule*[6pt]{.}\contentspage}
\titlecontents{subsubsection}[25pt]{\addvspace{5pt}\filright}              
{\contentspush{\thecontentslabel\ 
}}              
{}{\titlerule*[6pt]{.}\contentspage}
\begin{spacing}{1.2}
\tableofcontents
\end{spacing}
\newpage
\section{Introduction}
Heavy neutral gauge boson ($Z^{\prime}$) is predicted in various theories beyond the  Standard Model.  An attractive motivation came from the development of grand unified theory (GUT) with the extension of gauge group, such as those based on $SO(10)$ or $E_6$~\cite{Robinett:1981yz,London:1986dk,Langacker:1984dc,Hewett:1988xc,Kang:2004bz,Li:2022qrl,Lazarides:2019xai,Pernow:2019tuf,Benavides:2018fzm}. The models with extra dimensions also allow the  Kaluza-Klein excitations of gauge boson \cite{Han:1998sg,Duff:1986hr}.  Recently, the LHC has reported the research of  TeV scale $Z^{\prime}$ boson  via the process of dijet \cite{ATLAS:2019fgd,CMS:2019gwf}, dilepton \cite{ATLAS:2019erb,CMS:2021ctt}, diphoton \cite{ATLAS:2016gzy,CMS:2016kgr} , diboson \cite{ATLAS:2020fry,CMS:2021klu}  and $t \bar{t} $ \cite{ATLAS:2012dgv,CMS:2015fhb}. The model dependent excluded mass region with an upper bound can reach 4.5 (5.1) TeV for $Z_{\psi}$ $(Z_{SSM})$ boson~\cite{ATLAS:2019erb}. Meanwhile, the lepton colliders have the superiority of low background compared with the hadron colliders. It would be a better machine for the precision measurement on the extra neutral gauge boson at the future electron-positron colliders~\cite{Yin:2021rlr,CEPCStudyGroup:2018ghi}. Furthermore,  a muon collider, proposed with the less synchrotron radiation and higher collision energy, provides a bright prospect for the TeV $Z^\prime$ search. 

 A possible gauge group extension theory is to add an additional $U(1)$ group to the standard model group $SU(3)\times SU(2) \times U(1)$, which may result in models derived from GUT. Additional $U(1)$ group can also result from higher dimensional constructions such as string compactifications~\cite{Cvetic:1997mbh,Cvetic:1995rj}. In many models of GUT symmetry breaking, $U(1)$ groups survive at relatively low energies, resulting in corresponding neutral gauge bosons, commonly referred to as $Z^\prime$ bosons~\cite{Langacker:2008yv}.  Many other models predict the existence of $Z^\prime$ boson, such as 3-3-1 model~\cite{Cao:2016uur}, Minimal $Z^\prime$ model~\cite{Amrith:2018yfb,Deppisch:2018eth,Basso:2008iv} , Top-philic $Z^\prime$ model~\cite{Kim:2016plm,Greiner:2014qna} and Vector-leptoquark model~\cite{DiLuzio:2018zxy}. In these models, the coupling strength of $Z^\prime$ with standard model particles is different.  The properties of extra neutral gauge bosons have been extensively investigated in various models at the Large Electron-Positron Collider (LEP) and LHC~\cite{CDF:2008xbz,D0:2010kuq,Feldman:2006ce,Chiappetta:1996km,Barger:1996kr,CMS:2019buh,Rizzo:1985kn}. 
 
  Recently, a muon collider has been proposed with  great potential in high-energy physics. It can offer collisions of point-like particles at very high energies, since muons can be accelerated in a ring without limitation from synchrotron radiation~\cite{Delahaye:2019omf}. In the past few years, the International Muon Collider Collaboration (IMCC) has been actively exploring the construction of a muon collider with center-of-mass system (C.M.S.) energy of 10 TeV or higher and with high luminosity~\cite{Accettura:2023ked,Ruhdorfer:2023uea}. In our previous research, we have investigated the process of $Z'ZH$ interaction  at 14 TeV LHC and the future electron-positron collider~\cite{Li:2013ava,Yin:2021rlr}. An angular distribution  is suggested to  investigate  the  properties of $Z^\prime$ coupling to other particles. In this paper, we will  study the phenomenology of heavy neutral gauge boson at muon collider via the processes of $\mu^+\mu^- \rightarrow q \bar{q}$, $\mu^+\mu^- \rightarrow l^+ l^-$, $\mu^+\mu^- \rightarrow ZH$ and $\mu^+\mu^- \rightarrow W^+ W^-$. This study will provide detailed prospects for the heavy neutral gauge boson at the muon collider. 

This paper is organized as follows. In Section 2, we will give a detailed introduction to the theoretical framework. In Section 3, we study the scattering cross section and final state angular distribution of the processes with various $Z^\prime$ decay modes and the results with different beam polarizations are provided. Finally, a summary is presented.
\section{Theoretical Framework and Collider Constraints}
In order to introduce the $Z^\prime$ neutral gauge boson, we extend an additional $U(1)_X$ group in the gauge group structure of the standard model, $SU(3)_C \times SU(2)_L \times U(1)_Y$. We follow the notations in reference~\cite{Wells:2008xg} to depict the interactions and mass relationships in the Hidden Abelian Higgs Model (HAHM). The sector of this new group is coupled to the standard model particles only through kinetic mixing with the hypercharge gauge boson $B_\mu$. The kinetic energy terms of the $U(1)_X$ gauge group are
\begin{equation}\label{eq1}   
\mathcal{L}_K=-\frac{1}{4}\hat{X}_{\mu \nu} \hat{X}^{\mu \nu} + \frac{\chi}{2}\hat{X}_{\mu \nu}\hat{B}^{\mu \nu},
\end{equation}
where $\chi \ll 1$ is helpful to maintain the consistency between the accurate electric weak prediction and the experimental measurement results.
The gauge sector of Lagrangian can be written as
\begin{equation} \label{eq2}  
\mathcal{L}_G=-\frac{1}{4}\hat{B}_{\mu \nu} \hat{B}^{\mu \nu} - \frac{1}{4}\hat{W}_{\mu \nu}^a \hat{W}^{a \mu \nu} - \frac{1}{4}\hat{X}_{\mu \nu} \hat{X}^{\mu \nu}+\frac{\chi}{2}\hat{X}_{\mu \nu} \hat{B}^{\mu \nu},
\end{equation}
where $W^a_{\mu \nu}$, $B_{\mu \nu}$ and $X_{\mu \nu}$ are the field strength tensors of $SU(2)_L$, $U(1)_Y$ and $U(1)_X$ respectively. We diagonalize the $3 \times 3$ neutral gauge boson mass matrix and obtain the mass eigenstates as
\begin{equation}\label{eq3}   
\begin{pmatrix}
     B   \\
    W^3 \\
    X
\end{pmatrix}=
\begin{pmatrix}
     \cos\theta_W & -\sin\theta_W \cos\alpha & \sin\theta_W \sin\alpha   \\
     \sin\theta_W & \cos\theta_W \cos\alpha  & -\cos\theta_W \sin\alpha   \\
     0 & \sin\alpha & \cos\alpha
\end{pmatrix}
\begin{pmatrix}
     A   \\
    Z \\
    Z^\prime
\end{pmatrix}
,
\end{equation}
where the usual weak mixing angle and new gauge boson mixing angle are
\begin{equation}\label{eq4}
    \sin \theta_W = \frac{g^\prime}{\sqrt{g^2+g^{\prime 2}}},\qquad   \tan2\alpha=\frac{-2\eta \sin\theta_W}{1-\eta^2 \sin^2 \theta_W - \Delta z},
\end{equation}
with
\begin{equation}\label{eq5}
\Delta z = \frac{m_X^2}{m_{Z_0}^2},\qquad  m_{Z_0}^2 = \frac{(g^2+g^{\prime 2})v^2}{4},\qquad  \eta=\frac{\chi}{\sqrt{1-\chi^2}},
\end{equation}
where $m_{Z_0}$ and $m_X$ are the masses before mixing of $Z$ and $Z^\prime$. The mass eigenvalues of these two neutral gauge bosons are expressed as
\begin{equation}\label{eq6}
    m_{Z,Z^\prime}^2=\frac{m^2_{Z_0}}{2}\left[(1+\eta^2 \sin^2 \theta_W +\Delta z)\pm \sqrt{(1-\eta^2 \sin^2 \theta_W -\Delta z)^2+4\eta^2 \sin^2\theta_W} \right].
\end{equation}
From the above expression we can get  the mass eigenvalue  $m_Z \approx m_{Z_0}$  and $m_{Z^\prime}\approx m_X$ with $\eta$ in a suitable range. Therefore, The $Z^\prime ZH$ interaction can be parameterized as
\begin{equation}\label{eq7}
Z^\prime ZH : 2i \frac{m^2_{Z_0}}{v}(-\cos\alpha+\eta \sin\theta_W \sin\alpha)(\sin\alpha+\eta \sin\theta_W \cos\alpha).
\end{equation}
The $Z$ and $Z^\prime$ boson coupling to SM fermions can be expressed as
\begin{equation}\label{eq8}   
 \begin{split}
    \bar{\psi} \psi
    Z:&\frac{ig}{\cos\theta_W}[\cos\alpha(1-\eta \sin\theta_W \tan\alpha)]\gamma^{\mu} \left[T^3_L - \frac{(1-\eta \tan \alpha/ \sin \theta_W)}{(1-\eta \sin\theta_W \tan\alpha)}\sin^2\theta_W Q \right], \\
    &=-\frac{ig}{\cos\theta_W}\gamma^{\mu} \times \frac{1}{2}(c_V^{f}-c_A^{f} \gamma^5),
\end{split}
\end{equation}
\begin{equation}\label{eq9}   
\begin{split}
    \bar{\psi} \psi Z^\prime:&\frac{-ig}{\cos\theta_W}\left[\cos\alpha(\tan\alpha+\eta \sin\theta_W)\right]\gamma^{\mu} \left[T^3_L - \frac{(\tan\alpha+\eta/ \sin \theta_W)}{(\tan\alpha+\eta \sin\theta_W \tan\alpha)}\sin^2\theta_W Q \right] \\
    &=\frac{ig}{\cos\theta_W}\gamma^{\mu} \times \frac{1}{2}(c_V^{f\prime}-c_A^{f\prime} \gamma^5), 
\end{split}
\end{equation}
where $Q =T^3_L +Q_Y$, $T_L^3$ is the third component of  weak isospin, $Q_Y$ is hypercharge and $\theta_W$ is the Weinberg angle. $c_V^f$($c_V^{f\prime}$) and $c_A^f$($c_A^{f\prime})$ are the factors of vector and pseudovector component. We list the typical value of these parameters in Table \ref{tb1}  in  Appendix~\ref{appeA} with $\sin \alpha = 1\times 10^{-5}, 5 \times 10^{-5}$, and $1 \times 10^{-4}$, respectively.  

The mass value of $Z^\prime $ and $\eta$ are the main free parameters in the model. The mass region of $Z^\prime $ is predicted in the direct measurements from the lepton and hadron colliders~\cite{H1:2016goa,H1:2000bqr}. The mixing angle is determined by $\eta$ value from Equation~\eqref{eq3}, which is related to the mass of $Z$ from Equation~\eqref{eq6}. The accurate mass of $Z$ boson has been given experimentally with the value of $91.1876 \pm 0.0021$ GeV~\cite{ParticleDataGroup:2020ssz}. Thus, we choose the  $\eta$ and $\sin\alpha$ in a range that can guarantee the deviation of the $Z$ boson mass  in our model not beyond its experimental deviation. The decay width of $Z^\prime$ is taken  2\% of its mass as input in the following simulation of this paper.

$Z^\prime$ boson can come from more complicated modes such as GUT. In the context of most GUT models arising from the $E_6$ group, there exist two mixing angles, denoted as $\alpha$ and $\beta$. These mixing angles play a crucial role in the evolution process
 \begin{equation}
     E_6 \to SO(10)\times U(1)_{\psi} \to SU(5) \times U(1)_\chi \times U(1)_\psi.
 \end{equation}
 Specifically, $\beta$ represents the mixing angle between the $U(1)_\chi$ and $U(1)_\psi$ groups, whereby these two $U(1)$ groups combine linearly through the $\beta$ angle to form a novel $U(1)^\prime$. On the other hand, $\alpha$ is the mixing angle between $U(1)^\prime$ and the Standard Model's $U(1)_Y$. Various mixing angles give rise to distinct physical phenomena~\cite{Nandi:1986rg,Baer:1987eb,Barger:1987xw,Gunion:1987jd,Feldman:2006wb,Rizzo:2006nw,Lee:2008cn,Barger:2009xg,Gulov:2018zij}. Accordingly, different models have emerged based on these mixing angles~\cite{Erler:1999nx,Carena:2004xs,Erler:2002pr,Hicyilmaz:2022owb,Erler:2009jh}, such as the $U(1)_\chi$ model with an additional $U(1)\chi$~\cite{Robinett:1982tq}, the $Z_\psi$ model with an extra $U(1)_\psi$~\cite{Robinett:1982tq}, and the $Z_{B-L}$ model featuring a $U(1)_{B-L}$~\cite{Appelquist:2002mw}.
 
 Numerous experimental analyses from various colliders in the past have provided mass or coupling constraints on the possible existence of $Z^\prime$ particles within different models. The simple extension to SM $Z$ boson is the so-called  Sequential Standard Model (SSM) which implies the existence of  $Z^\prime$ with the same couplings as the $Z$ boson except the mass value. The CMS collaboration has determined the lower mass limits for  the Sequential Standard Model $Z_{SSM}$ in the channels $Z^\prime \to e^+ e^-$, $Z^\prime \to \mu^+\mu^+$, and $Z^\prime \to e^+ e^-$ or $\mu^+\mu^-$ to be 4100 GeV, 4250 GeV, and 4500 GeV, respectively, based on the analysis of proton-proton collision data at $\sqrt{s} = 13$ TeV with an integrated luminosity of $36.1$ $\text{fb}^{-1}$. The mass lower limits for the three decay channels corresponding to the $Z^\prime_\psi$ model are 3450 GeV,  3700 GeV, and 3900 GeV~\cite{CMS:2018ipm}, respectively. The ATLAS experiment has established exclusion limits at a $95\%$ confidence level for SSM $Z^\prime$ bosons decaying into a pair of $\tau-$leptons, with a lower mass limit of   2420  GeV. Additionally, for the non-universal $G(221)$ model, which demonstrates enhanced couplings to third-generation fermions, the exclusion limit is set at  2250 GeV~\cite{ATLAS:2017eiz}.  Studies have been conducted for high-mass resonances in the dijet invariant mass spectrum. In the context of the SSM, $Z^\prime$ gauge bosons with $Z^\prime \to b \bar{b}$ decays are excluded for masses up to $2000 \text{GeV}$, and in the leptophobic model with Standard Model-like couplings to quarks, $Z^\prime$ gauge bosons are excluded for masses up to 2100 GeV at a $95\%$ confidence level~\cite{ATLAS:2018tfk} . Several prominent models and the current experimental lower mass limits are presented in Table~\ref{otherModel} for reference.
 \begin{table}[H]
\begin{center}
\setlength{\abovecaptionskip}{6pt}
\setlength{\belowcaptionskip}{0pt}
\centering
\begin{tabular}{c c c c c}
\toprule
$Z^\prime$ model & $ \quad \alpha$ \quad &  $ \quad \beta$ \quad  &Group&Lower Limit of $M_{Z^\prime}$ \\
\midrule
$Z^\prime_\chi$ &\quad$0^\circ$&\quad$0^\circ$ &$SO(10)\to SU(5)\times U(1)_\chi$ &4100 GeV  \\
$Z^\prime_\psi$ &\quad$0^\circ$&\quad$90^\circ$ &$E_6 \to SO(10)\times U(1)_\psi$ &3900 GeV \\
$Z^\prime_\eta$ &\quad$0^\circ$&\quad$-52.2^\circ$ &$E_6 \to SU(3)\times SU(2) \times U(1) \times U(1)_\eta$&3900 GeV  \\
$Z^\prime_{LR}$ &\quad$-39.2^\circ$&\quad$0^\circ$ &$SU(2)_L\times SU(2)_R\times U(1)$ &1162 GeV \\
\bottomrule
\end{tabular}
\caption{The mass constraints for various models including $Z^\prime$ at the LHC~\cite{ParticleDataGroup:2020ssz}.}
\label{otherModel}
\end{center}
\end{table}
 For a hadron collider, the observational limits can be provided through a model-independent analysis. If one denotes by $N_{Z^\prime}$ the number of signal events required for a discovery signal,  the upper limit on the  mass of $X$ is defined as~\cite{Leike:1997cw},
\begin{equation}
  m_{Z^\prime}^{lim} \approx \frac{\sqrt{s}}{A}\ln \left(\frac{L}{s}\frac{c_{Z^\prime} C}{N_{Z^\prime}}\right),
\end{equation}
where the details of the model are encoded in 
\begin{equation}
  c_{Z^\prime}=\frac{4 \pi^2}{3}\frac{\Gamma_{Z^\prime}}{m_{Z^\prime}}B(\mu^+ \mu^-)\left[B(u \bar{u})+\frac{1}{C_{ud}B(d\bar{d})}\right]
\end{equation} 
For a $pp(p\bar{p})$ collider, $A=32~(20)$, $C=600~(300)$, and in the kinematical region of interest at LHC $C_{ud}\sim 2$. The predicted upper limit for $Z^\prime$ is around 5000 GeV at a Large Hadron Collider with $\sqrt{s}=14$ TeV~\cite{Leike:1997cw}.  Furthermore, the coupling constant $g^{\prime}_{q}$ for $Z^\prime$ coupling to $q\bar{q}$ has been introduced in some models. 
 Experimental groups like ATLAS~\cite{ATLAS:2018hbc,ATLAS:2019itm,ATLAS:2018qto,ATLAS:2019fgd}, CMS~\cite{CMS:2018mgb,CMS:2018ucw,CMS:2016ltu,CMS:2019gwf,CMS:2019mcu,CMS:2018kcg,CMS:2018mgb,CMS:2019emo,CMS:2019xai,CMS:2018rkg}, CDF~\cite{CDF:2008ieg,CDF:1997chq} and UA2~\cite{UA2:1993tee} have provided constraints on $g^{\prime}_q$ in various decay channels. Most of these exclusion lower limitations are  the order of $0.1$, and the most stringent limit coming from the study of dijet process by ATLAS with a lower bound of 0.05 when $M_{Z^\prime}\sim 2$ TeV~\cite{ATLAS:2018hbc}.

 Studies of the $Z^\prime$ boson have also been conducted at lepton colliders, particularly in the low mass regions below 1000 GeV. Experiments carried out by ALEPH and OPAL at LEP have constrained the mass of the $Z^\prime$ boson through the $e^+ e^- \to  f f$ process, where $f$ represents either a $\tau$ lepton or a $b$ quark. The lower mass limits for the $Z^\prime$, derived from $\tau-$pair production, are  365 GeV for ALEPH and 355 GeV for OPAL, while those obtained from $b \bar{b}$ production are 523 GeV for ALEPH and 375 GeV  for OPAL~\cite{Lynch:2000md}. With the prospect of future lepton colliders, there is potential to significantly extend the mass range and coupling strength of the $Z^\prime$ boson, especially in the context of Higgs boson production.

For the Hidden Abelian Higgs Model, the new gauge sector couples with the Standard Model solely through the dynamical mixing with the hypercharge gauge boson, denoted as $B_\mu$. Many experiments are more concerned with the coupling constraints between $Z^\prime$ and dark matter particles, with $Z^\prime$ typically taking values smaller than the mass of the $Z$ boson~\cite{ATLAS:2018coo,Diamond:2017ohe,Liu:2017jzn}. Some experimental and phenomenological studies have also imposed constraints on this model in other parameter spaces~\cite{Curtin:2014cca,ATLAS:2021ldb}.   The limitations on the $\eta$ value in the Hidden Abelian Higgs Model are investigated in different colliders. For example, it has been suggested that there maybe an opportunity to detect signals of the $Z^\prime$ particle in this model when $\eta$ exceeds 0.03 at the $\sqrt{s}=13~\text{TeV}$ LHC with the integrated luminosity of 100 $\text{fb}^{-1}$ in the mass region of $600\sim1400$ GeV. For the International Linear Collider (ILC)  at $ \sqrt{s} = 500~\text{GeV}$ for 500 $\text{fb}^{-1}$ integrated luminosity, a $Z^\prime$ mass of $750~(1000)~\text{GeV}$ requires $\eta$ value greater than $0.1~(0.15)$ in order to achieve a significance level exceeding $5\sigma$  in the process $e^+ e^- \to \mu^+ \mu^-$~\cite{Kumar:2006gm}.

\section{ $Z^\prime$ Boson Production and Decay at Future Muon Collider}
\begin{figure}[H] 
\centering
\includegraphics[width=8cm,height=6cm]
{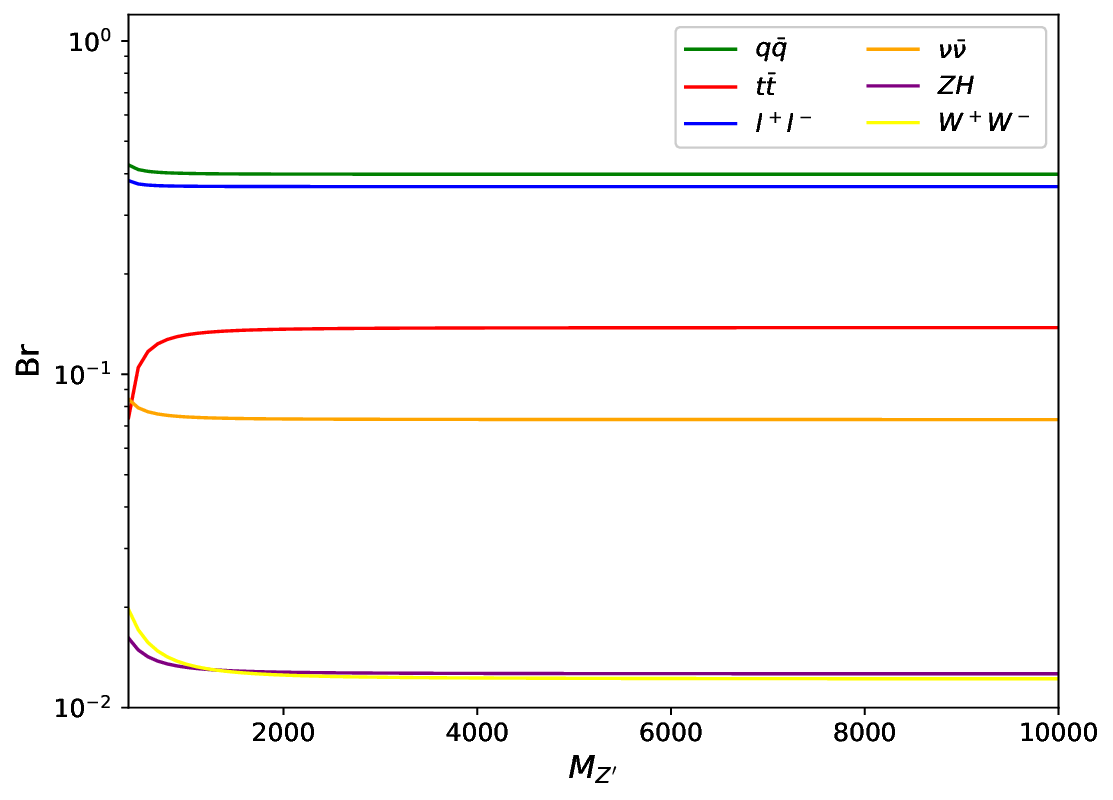}
\caption{$Z^\prime$ decay branching ratio.}
\label{Br}
\end{figure}

In this part, we will study the $Z^\prime$ production and decay  with the process of  $\mu^+ \mu^- \rightarrow q \bar{q}$, $\mu^+ \mu^- \rightarrow l^+ l^-$, $\mu^+ \mu^- \rightarrow Z H$ and $\mu^+ \mu^- \rightarrow W^+ W^-$. The main decay branch ratios of  $Z^\prime$ are illustrated in Figure~\ref{Br}.  $Z^\prime \to q \bar{q}$ and $Z^\prime \to l^+ l^-$ are the dominant decay modes with the branching ratio around 40\% and 35\%. The branching ratio of $t \bar t$ can reach 14\% with the mass of $Z^\prime$ larger than the threshold of top pair production. Although the branching ratio of  $Z^\prime \rightarrow W^+ W^-$ and $Z^\prime \rightarrow ZH$ is only about 1\%, it is useful to study these decay processes for the interaction between $Z^\prime$  and other gauge bosons and Higgs particle. The decay modes of $Z^\prime \rightarrow \nu \bar{\nu}$ is also an interesting process for the invisible decay, which is out of the reach of this paper.

\begin{figure}[H] 
	\center
\includegraphics[width=16cm,height=6cm]{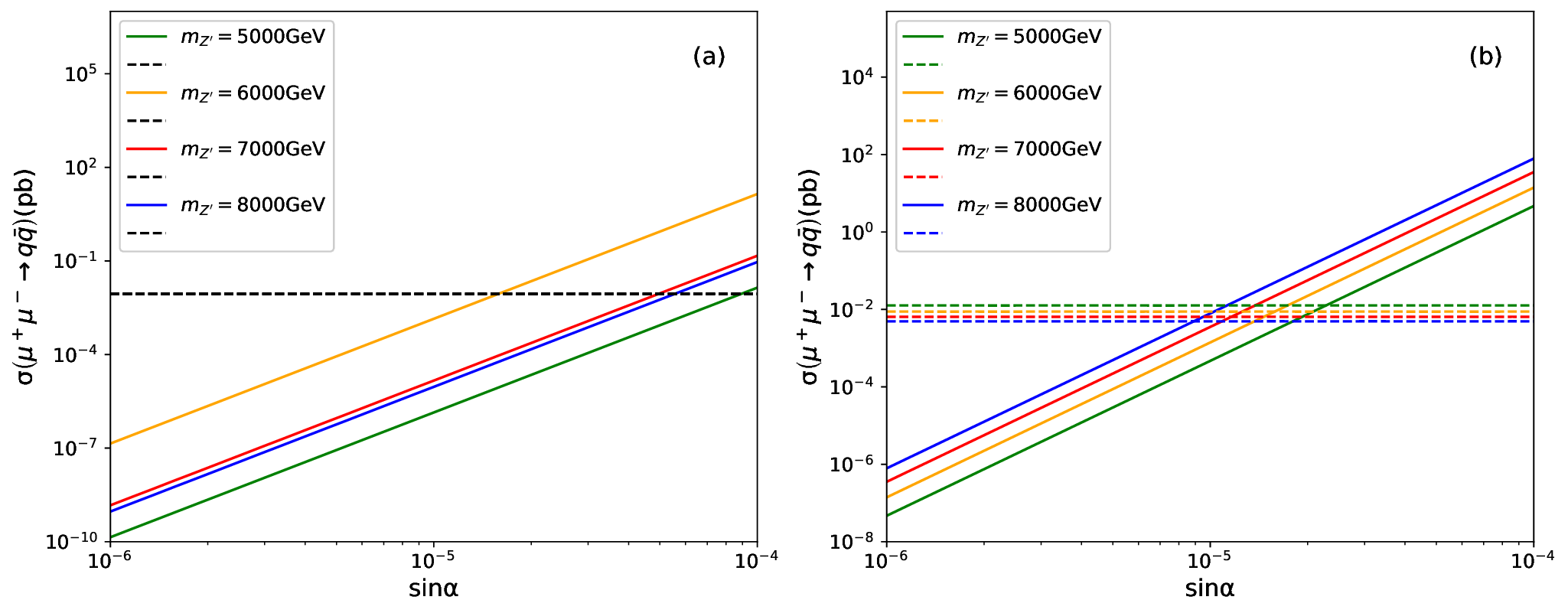}
 \setlength{\abovecaptionskip}{0cm}
 \caption{The cross section of $\mu^+ \mu^- \rightarrow Z \rightarrow q \bar{q}$ (dashed) and $\mu^+ \mu^- \rightarrow Z^\prime \rightarrow q \bar{q}$ (soild) versus $\sin\alpha$ with $\sqrt{s}=6000$ GeV (a) and $\sqrt{s}=m_{Z^\prime}$ (b) respectively.   }
  \label{etaAndSigmaMumuToqq}
\end{figure}

We give an example for the production of $\mu^+ \mu^- \rightarrow Z^\prime \rightarrow q \bar{q}$ at the muon collider. The cross section distribution as a function of $\sin\alpha$ is shown in  Figure~\ref{etaAndSigmaMumuToqq}.  Comparing  with the result of $\mu^+ \mu^- \rightarrow Z \rightarrow q \bar{q}$,  the  cross section induced by $Z^\prime$ increases with the increasing  of $\sin \alpha$.  The  values of the  vertex factor in Equations~\eqref{eq7} and \eqref{eq8} with the typical value of  $\sin \alpha$ are presented  in Table~\ref{tb1}.  

\begin{figure}[H] 
\centering
\includegraphics[width=8cm,height=6cm] 
{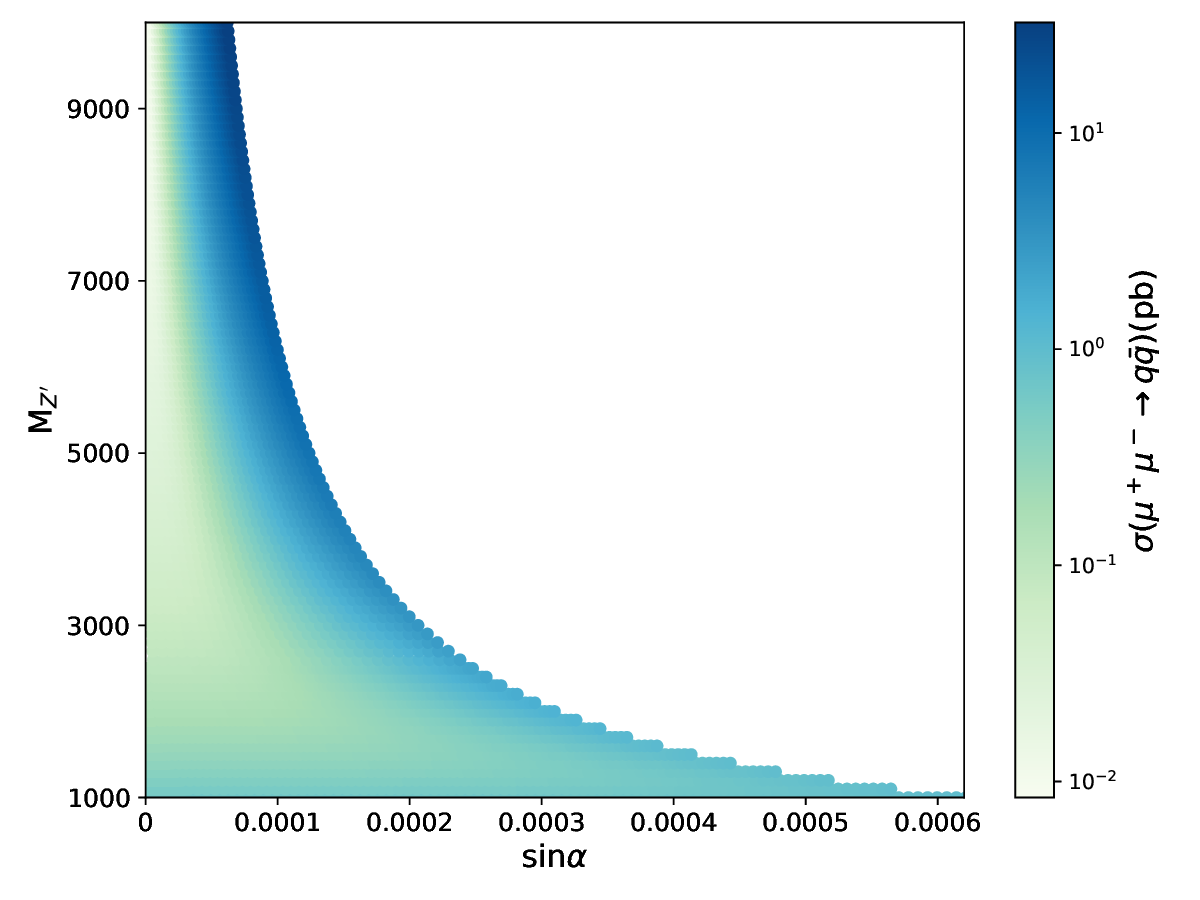}
\caption{ Contour map of cross section for process $\mu^+ \mu^- \rightarrow \gamma^*/Z/Z^\prime \rightarrow q \bar{q}$ on the $\sin\alpha-M_{Z^\prime}$ plane with $\sqrt{s}=M_{Z^\prime}$.}
\label{qqdifAlphaZp}
\end{figure}

In Figure~\ref{etaAndSigmaMumuToqq} (a), the cross section induced by $Z$ boson is 0.0088 pb with $\sqrt{s}=6000$ GeV, and it will be the same value in the $Z^\prime$ induced process when $m_{Z^\prime}=6000$ GeV and $\sin\alpha=1.6 \times 10^{-5}$.   The distribution of  cross section has the maximum value with $m_{Z^\prime}=6000$ GeV for the resonance enhancement with  the C.M.S. energy close to  $m_{Z^\prime}$. While the cross sections show different orders with various mass values  in Figure~\ref{etaAndSigmaMumuToqq} (b). In this case the cross section increases with the increasing of  $m_{Z^\prime}$ where the resonance  enhancement is included in each case for $m_{Z^\prime}=\sqrt{s}$.  The cross sections of other processes induced by $Z^\prime$ boson, such as  $\mu^+ \mu^- \rightarrow l^+ l^-$, $\mu^+ \mu^- \rightarrow Z H$ and $\mu^+ \mu^- \rightarrow W^+ W^-$, have the same distributions which can be obtained using the proportions of the decay branching ratio. In order to  directly display the dependence between  the mass of $Z^\prime$ boson, the mixing angle $\sin\alpha$ and the production of $Z^\prime$, we give the cross section of $\mu^+ \mu^- \rightarrow \gamma^*/Z/Z^\prime \rightarrow q \bar{q}$ projected to the $\sin\alpha-M_{Z^\prime}$ plane  in Figure~\ref{qqdifAlphaZp}. In this plot we have considered the constraint on $\sin\alpha$ from the measurement of $Z$ boson mass. We have required that the mass correction of $Z$ boson according to Equation~\eqref{eq6} should be less than the experimental deviation in~\cite{ParticleDataGroup:2020ssz},  thus the blank space in Figure~\ref{qqdifAlphaZp} is excluded by the experimental measurement. The cross section is  insensitive to $\sin \alpha$ when the mass of $Z^\prime$ is relatively small. 
 As $m_{Z^\prime}$ increases, the allowed range of  $\sin\alpha$ decreases, meanwhile the scattering cross section gradually becomes sensitive to $\sin \alpha$. 
 This means that when  $Z^\prime$ mass is large, a suitable $\sin \alpha$ can lead to a larger cross section. With $m_{Z^\prime}=6000$ GeV and  $\sin\alpha \approx 1\times 10^{-4}$, the cross section can be reached 12 pb at the C.M.S. energy of 6 TeV. In this figure, the maximum cross section value is about 32 pb, which corresponds to $m_{Z^\prime}= 10000$ GeV and $\sin\alpha =6 \times 10^{-5}$. Similar to Figure~\ref{etaAndSigmaMumuToqq}, the cross section of other processes induced by $m_{Z^\prime}$ have the similar dependence on  $m_{Z^\prime}$ and $\sin \alpha$. We will study the $Z^\prime$ effects in each decay channel in the following sections. 
\subsection{$Z^\prime$ search in  $\mu^+ \mu^- \rightarrow q \bar{q}$}
{\subsubsection{The cross section of process $ \mu^+  \mu^- \rightarrow q \overline{q}$}}
 \begin{figure}[tbph] 
\centering
\setlength{\belowcaptionskip}{-2mm}
\includegraphics[width=16cm,height=18cm]{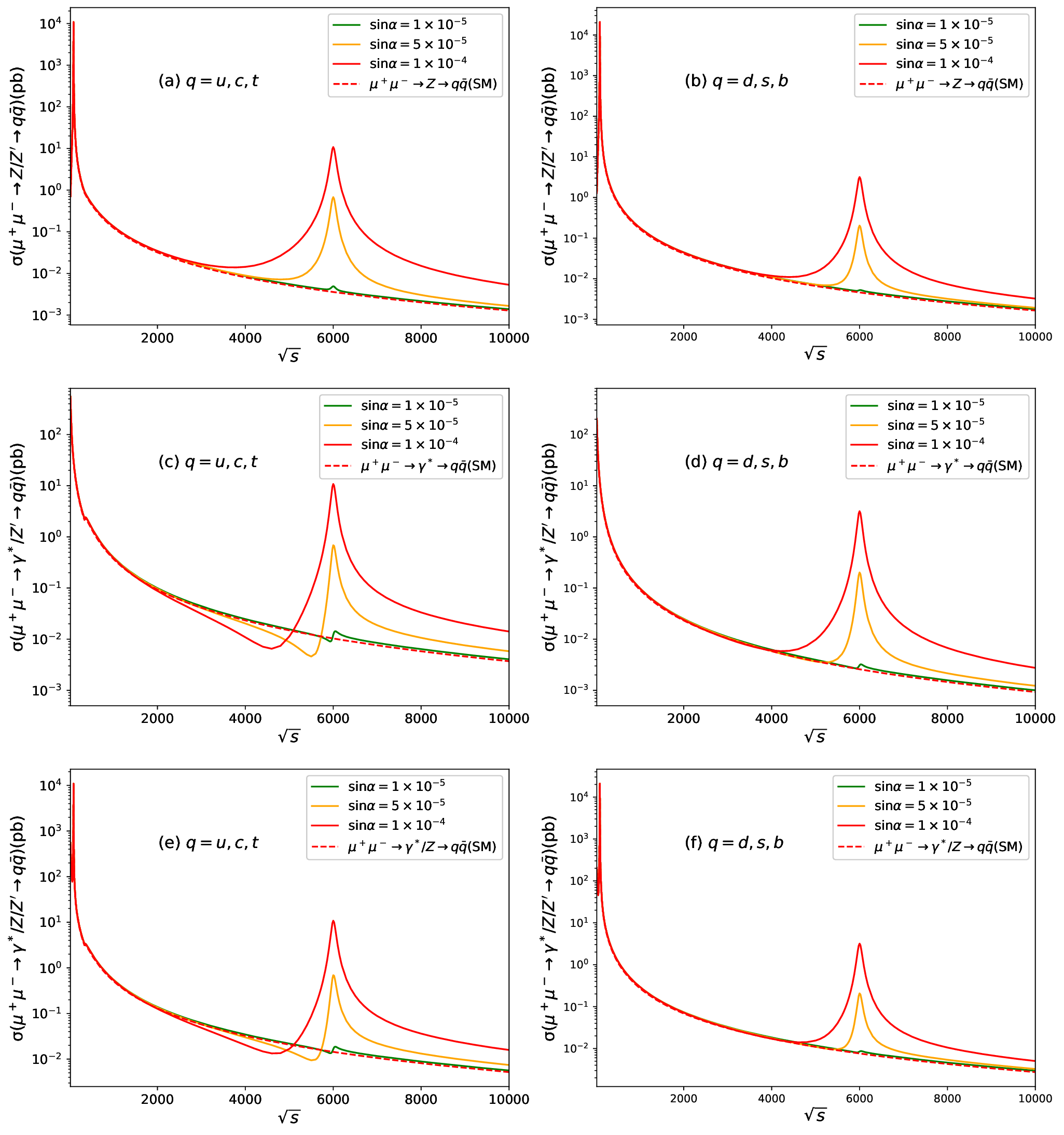}
\caption{The cross section of process $\mu^+ \mu^- \rightarrow  q \overline{q}$ versus $\sqrt{s}$ with $m_{z^\prime}=6000$ GeV. The final particles are $q \bar{q}$ where $q=u,c,t$ in (a), (c), (e) and $q = d, s, b$ in (b),(d),(f).}
 \label{CrossMuMuToqq}
\end{figure}
We have simulated the production of quark final state processes with all flavors. Through the analysis of the results, we  divide the quark processes into two categories according to the electrical properties of quarks, where $u, c, t$ are quarks with positive charges, and $d, s, b$ are quarks with negative charges. In Figure~\ref{CrossMuMuToqq}, we show the variation of the scattering cross sections of these two kinds of quark final state processes with different C.M.S. energy. When the C.M.S. energy approaches the mass of $Z^\prime$, a prominent resonance peak will manifest, which is consistent with the phenomenon observed in Figure~\ref{etaAndSigmaMumuToqq}. Another interesting phenomenon occurs in the $\mu^+ \mu^- \rightarrow Z^\prime/\gamma^* \rightarrow q \bar{q}$ process. If the final state quarks carry positive charge, the scattering cross section will exhibit  suppressions with different $\sin\alpha$ values as the C.M.S. energy approaching and falling below the $Z^\prime$ mass. However, this phenomenon is not observed when the final state quarks carry negative charge. The discrepancy is obvious in Figure~\ref{CrossMuMuToqq} (e) and (f), which is due to the negative interference effect of $Z^\prime$ and $Z, \gamma^*$. The detailed analysis can be found in Appendix~\ref{appeA}. Furthermore, the suppression of the scattering cross section increases with the increasing value of $\sin \alpha$ and a larger resonance peak appears as shown in Figure~\ref{CrossMuMuToqq}.
{\subsubsection{The angular distribution of quarks in $\mu^+ \mu^- \rightarrow q \bar{q}$}}
As we all know, the top quark will decay before hadronization which will not be the same as other quarks. We firstly focus on investigating processes involving  light quarks excluded the  top quark.  Following the discussion in reference~\cite{Lynch:2000md}, we study the interaction between $Z^\prime$ and other particles by analyzing the angular distribution of the final particles. The angle can be expressed from the following formula:
\begin{equation}\label{eq13}
    \cos\theta=\frac{\bm{p_{f}^*} \cdot \bm{p_{i}}}{|\bm{p_{f}^*}|\cdot |\bm{p_{i}} |},
\end{equation}
where $\bm{p_{f}^*} $ and $\bm{p_{i}}$ are the three momentum of the final and initial particle, respectively. \par
\begin{figure}[tbh] 
\center
	\includegraphics[width=16cm,height=5.5cm]{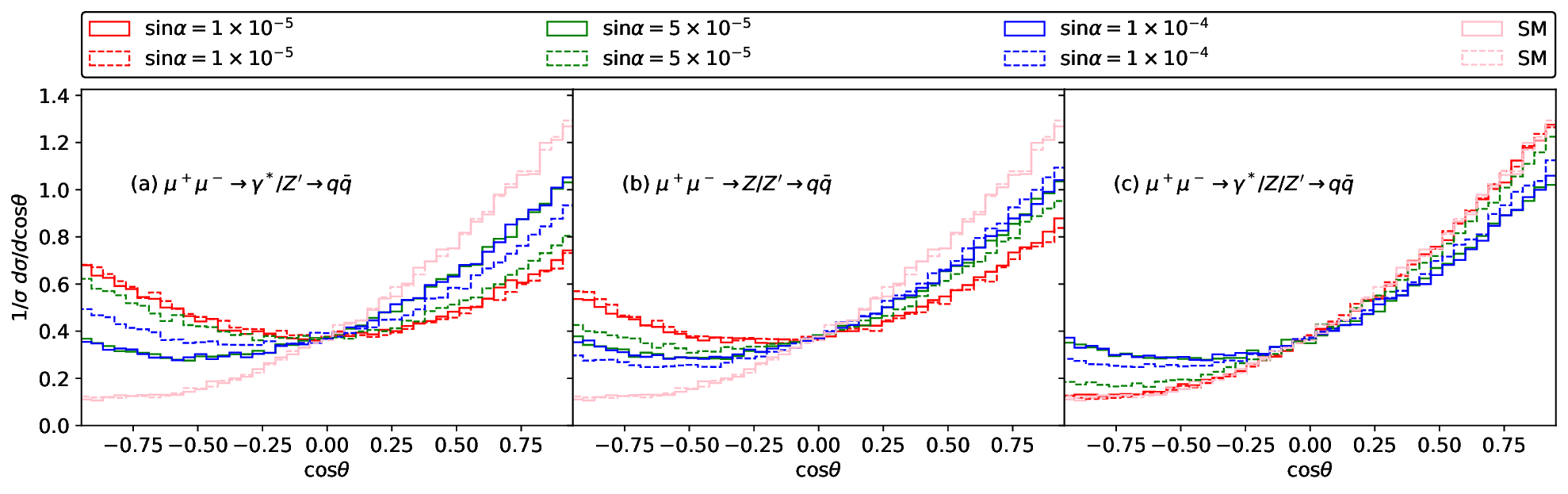}
	\caption{Angular distribution of  final particles of process $\mu^+ \mu^- \rightarrow \gamma^* /Z^\prime \rightarrow q \bar{q}$ (a), $\mu^+ \mu^- \rightarrow Z /Z^\prime \rightarrow q \bar{q} $ (b) and $\mu^+ \mu^- \rightarrow \gamma^*/Z/Z^\prime \rightarrow q \bar{q} $ (c) with $m_{Z^\prime}=6000$ GeV and the $\sqrt{s}$ is 6000 GeV(solid) and 8000 GeV(dashed).}
    \label{qqDistribution}
\end{figure}
\begin{figure}[H] 

\centering
\includegraphics[width=7cm,height=4cm]
{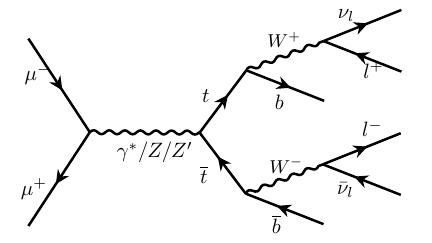}
\caption{Feynman diagram of $\mu^+ \mu^- \rightarrow t \bar{t} \rightarrow b \bar{b} W^+ W^- \rightarrow l^+ l^- \nu_l \bar{\nu}_l$.}
\label{ttDecayFeyDia}
\end{figure}
\begin{figure}[tbh] 
\center
	\includegraphics[width=16cm,height=5.5cm]{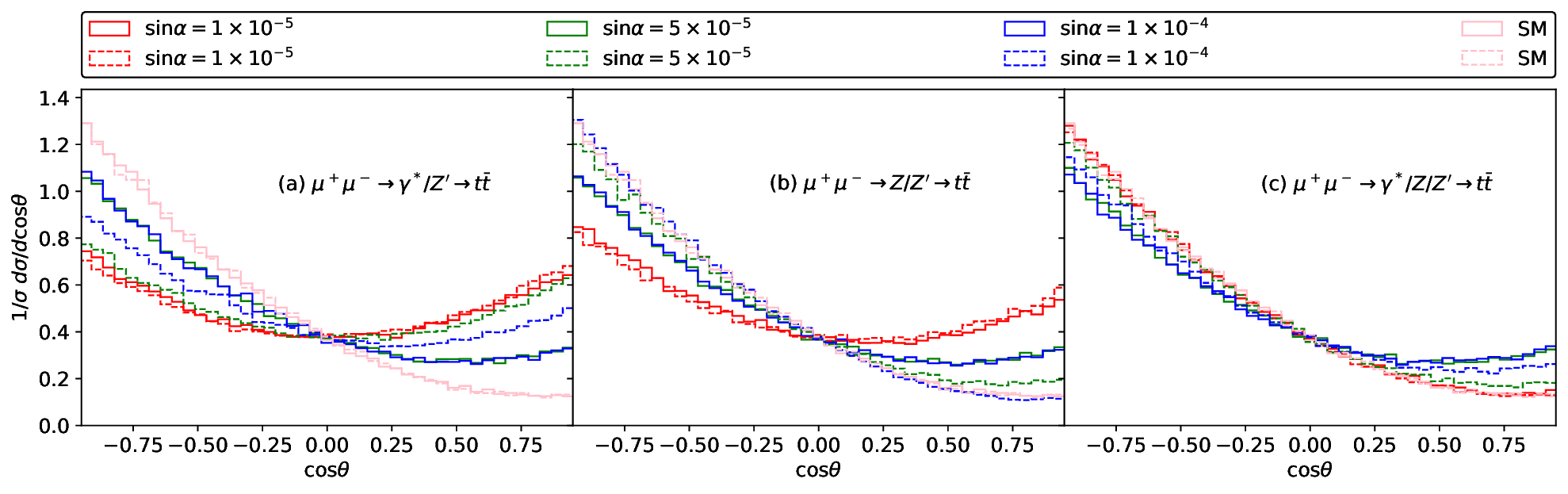}
	\caption{Angular distribution of final leptons of process $\mu^+ \mu^- \rightarrow \gamma^* /Z^\prime \rightarrow t \bar{t} \rightarrow b \bar{b} l^+ l^- \nu_l \bar{\nu}_l$ (a), $\mu^+ \mu^- \rightarrow Z/Z^\prime \rightarrow t \bar{t} \rightarrow b \bar{b} l^+ l^- \nu_l \bar{\nu}_l$ (b) and $\mu^+ \mu^- \rightarrow\gamma^*/ Z/Z^\prime \rightarrow t \bar{t} \rightarrow b \bar{b} l^+ l^- \nu_l \bar{\nu}_l$ (c) ($l^+ l^-$ possible combinations are $\mu^+ \mu^-, e^+ e^-, e^+ \mu^-, e^- \mu^+$) with $m_{Z^\prime}=6000$ GeV and C.M.S. energy is 6000 GeV (solid) and 8000 GeV (dashed).}
    \label{ttDistribution}
\end{figure}

Figure~\ref{qqDistribution} shows the angular distribution for the final quarks $q (q=u,c,d,s,b)$ with $m_{Z^\prime}=6$ TeV and the C.M.S. energy $\sqrt{s}=6$ TeV and 8 TeV. We present the final quark angular distributions for $\gamma^*/Z^\prime $, $Z/Z^\prime$ and $\gamma^*/Z/Z^\prime$ processes respectively, from which a highly asymmetric angular distribution of the final state particles can be obtained. With the increasing of $\sin\alpha$, the angular distributions with $Z^\prime$ contribution deviate far from the SM distribution in Figure~\ref{qqDistribution} (c). 
A forward-backward  asymmetry according to the angular distribution can be defined as
\begin{equation}\label{eq14}
    A_{FB}=\frac{\sigma(\cos\theta \geq 0)-\sigma( \cos \theta <0)}{\sigma(\cos\theta \geq 0)+\sigma( \cos \theta <0)}.
\end{equation}
The corresponding forward-backward asymmetries of $\mu^+ \mu^- \rightarrow q \bar{q}$ process are listed in Table \ref{FBAmumuToqq}. The asymmetry becomes large with the C.M.S. energy changing from 6000 GeV to 8000 GeV. With the increasing of $\sin\alpha$, the asymmetry changes small and far from the SM. 
\begin{table}[H]
\begin{center}
\setlength{\abovecaptionskip}{6pt}
\setlength{\belowcaptionskip}{0pt}
\begin{tabular}{ c c c c c c c c}
\toprule
 $\sqrt{s}$\qquad & $\sin \alpha$\qquad & $A_{FB}(Z/Z^\prime)$\qquad & $A_{FB}(\gamma^*/Z^\prime)$\qquad & $A_{FB}(\gamma^*/Z/Z^\prime)$ \\
\midrule
\multirow{3}*{6000 GeV} \qquad & $1\times 10^{-5}$\qquad & $0.171$\qquad  & $0.034$\qquad & $0.612$\\
~ & $5\times 10^{-5}$\qquad &  $0.358$\qquad  &  $0.360$\qquad & $0.377$\\
~ & $1\times 10^{-4}$\qquad & $0.366$\qquad  & $0.366$\qquad & $0.373$\\
 & & & & \\
 \multirow{3}*{8000 GeV} \qquad & $1\times 10^{-5}$\qquad &  $0.143$\qquad  &  $0.005$\qquad & $0.624$\\
~ & $5\times 10^{-5}$\qquad & $0.289$\qquad  & $0.093$\qquad & $0.554$\\
~ & $1\times 10^{-4}$\qquad & $0.432$\qquad  & $0.226$\qquad & $0.443$\\
\bottomrule
\end{tabular}
\caption{Forward-backward asymmetry for the process $\mu^+ \mu^- \rightarrow q \bar{q}$ with $m_{Z^\prime}=6000$ GeV.}
\label{FBAmumuToqq}
\end{center}
\end{table}
Due to the short lifetime of $t$ quark, we will discuss $t$ quark processes separately. The leptonic decay modes for $t$ quarks via $t \bar{t} \rightarrow b \bar{b} W^+ W^- \rightarrow b  \bar{b} l^+ l^- \nu_l \bar{\nu}_l(l=e, \mu)$ is selected in our study. Figure~\ref{ttDecayFeyDia} displays the corresponding Feynman diagram. The angular distributions of the final leptons are investigated for studying   $\mu^+ \mu^- \rightarrow t \bar{t}$ in the future muon colliders with the cascade decay  shown in Figure~\ref{ttDistribution}. The angular distribution of the final leptons is utilized to study the $Z^\prime$ properties in the top pair production, where the final lepton angular distribution $\cos\theta$ is defined as formula \eqref{eq13} with $\bm{p_{f}^*} $ being $\bm{p_{l^-}^*} $ and $\bm{p_{i}}$ being $\bm{p_{\mu^-}}$, respectively. It is noteworthy that the dilepton signal produced by the decay channel of $t$ quarks differs significantly from that produced by the direct decay of $Z$ or $Z^\prime$ bosons to diquarks. 

The major difference  derives from the following two parts. Firstly, the dilepton directly generated by $Z$ or $Z^\prime$ can only be the same generation of leptons to meet the conservation of lepton numbers, namely $e^+ e^-$ and $\mu^+ \mu^-$ (where $\tau^+\tau^-$ is excluded for its short life time and different detection methods). But the dilepton produced by the decay channel of $t$ quark actually decay from $W^+$ and $W^-$ to leptons and corresponding neutrinos, so dilepton can be the combination of $e^+e^-$, $\mu^+ \mu^-$, $e^- \mu^+$, $e^+ \mu^-$.  Secondly, since the final state particle we choose is a particle rather than an antiparticle,  the  angular distribution in the $t\bar{t}$ process should be inverse of the  light quark process, which is caused by $l^-$ coming from the decay of $\bar{t}$ quark.  The  forward-backward asymmetry of the angular distribution for the $t\bar{t}$ process are listed in Table \ref{FBAmumuTottDecay}. With the same value of $m_{Z^\prime}$, as the mixing angle increases, the angular distribution deviates from the standard model further. The asymmetry will be slightly enhanced with the C.M.S. energy increasing in the process of intermediate $\gamma^*/ Z/Z^\prime$ as shown in  Figure~\ref{ttDistribution} (c).

\begin{table}[H]
\begin{center}
\setlength{\abovecaptionskip}{6pt}
\setlength{\belowcaptionskip}{0pt}
\begin{tabular}{ c c c c c }
\toprule
 $\sqrt{s}$\qquad & $\sin \alpha$\qquad & $A_{FB}(Z/Z^\prime)$\qquad & $A_{FB}(\gamma^*/Z^\prime)$\qquad & $A_{FB}(\gamma^*/Z/Z^\prime)$ \\
\midrule
\multirow{3}*{6000 GeV} \qquad & $1\times 10^{-5}$\qquad & $-0.174$\qquad  & $-0.036$\qquad & $-0.594$\\
~ & $5\times 10^{-5}$\qquad &  $-0.398$\qquad  &  $-0.389$\qquad & $-0.399$\\
~ & $1\times 10^{-4}$\qquad & $-0.404$\qquad  & $-0.398$\qquad & $-0.395$\\
 & & & & \\
 \multirow{3}*{8000 GeV} \qquad & $1\times 10^{-5}$\qquad &  $-0.136$\qquad  &  $-0.003$\qquad & $-0.609$\\
~ & $5\times 10^{-5}$\qquad & $-0.544$\qquad  & $-0.078$\qquad & $-0.554$\\
~ & $1\times 10^{-4}$\qquad & $-0.629$\qquad  & $-0.217$\qquad & $-0.466$\\
\bottomrule
\end{tabular}
\caption{Forward-backward asymmetry for the process $\mu^+ \mu^- \rightarrow t \bar{t} \rightarrow b\bar{b}l^+ l^- \nu_l \bar{\nu}_l$ with $m_{Z^\prime}=6000$ GeV.}
\label{FBAmumuTottDecay}
\end{center}
\end{table}
{\subsubsection{Signals with beam polarization}}

\begin{figure}[tbh] 
\centering
\includegraphics[width=8cm,height=6cm]
{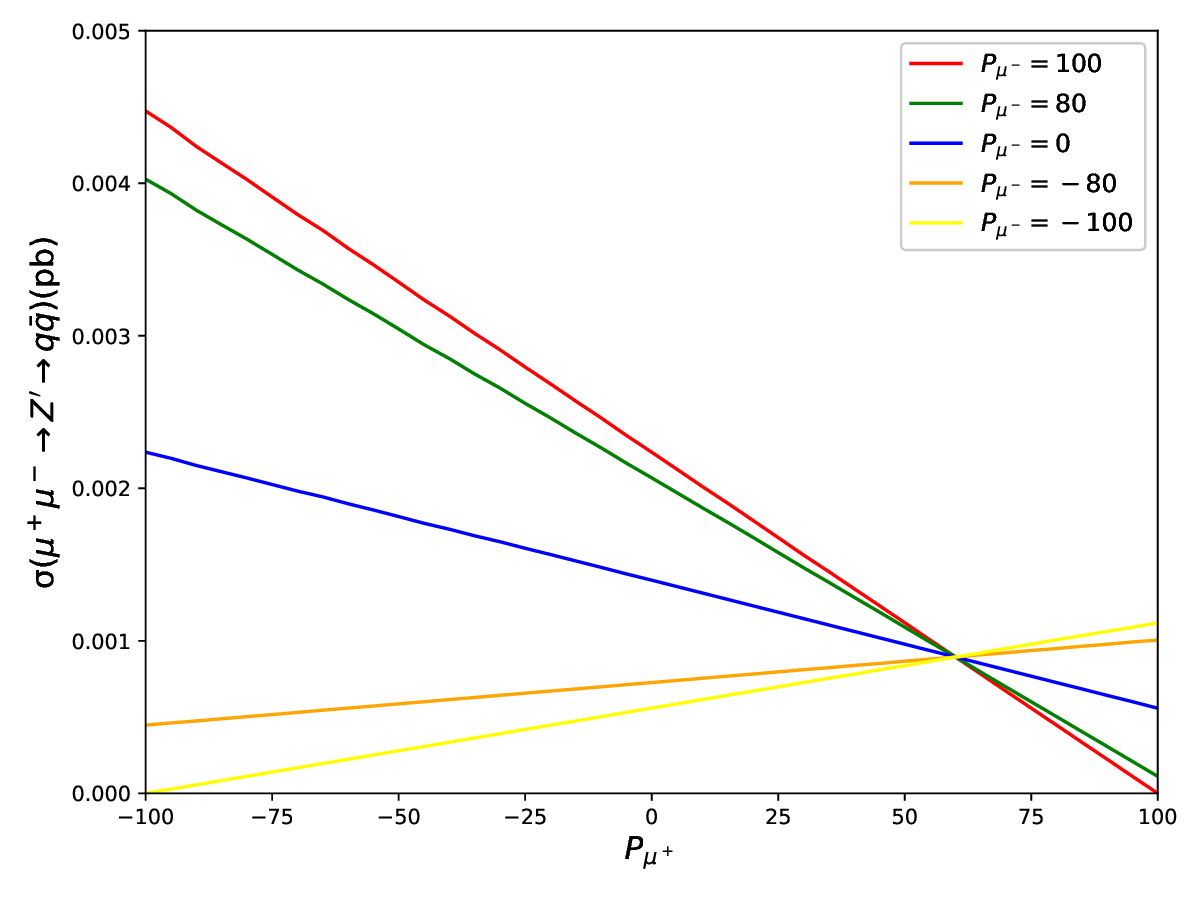}
\caption{The cross section of $\mu^+ \mu^- \rightarrow Z^\prime \rightarrow q \bar{q}$ with polarized $\mu^-$ and $\mu^+$ beam for $M_{Z^\prime}=6000$ GeV and $\sin \alpha = 1 \times 10^{-5}$.}
\label{crosOfpolarized}
\end{figure}

The beam polarization has superiority at the lepton colliders, which can be employed to enhance the effects from $Z^\prime$ boson in some ways. In the case of the Circular Electron-Positron Collider (CEPC),  the choice has been made to employ a polarized electron source. In order to facilitate the future collision between polarized electron beams and unpolarized positron beams, the Circular Electron Positron Collider (CEPC) contemplates the utilization of a specially prepared GaAs/GaAsP superlattice photocathode DC gun-type electron source. This implementation is expected to result in the production of polarized electron beams with a polarization degree exceeding $85\%$ on average~\cite{CEPCStudyGroup:2018rmc}. Although it is not easy to keep the large luminosity associated with the large beam polarizations stated by J. F. Gunion~\cite{Gunion:1998bc}, 
 the prospects of muon beam polarization on future circular colliders have also been widely discussed~\cite{Accettura:2023ked,AlAli:2021let,Liu:2023yrb,Belfkir:2023lot,Black:2022cth}, with the International Muon Collider Collaboration presenting exclusion regions for luminosity at different collision energies when muon beam polarization is set at $-30\% (+30\%)$~\cite{MuonCollider:2022xlm}.We investigate the $\mu^+ \mu^- \rightarrow Z^\prime \rightarrow q \bar{q}$ process with different polarizations of muon. The cross section is shown in Figure~\ref{crosOfpolarized} with the polarization of $P_{\mu^-}= 100, 80, 0, -80, -100$ and $P_{\mu^+}$ values ranging from $-100$ to $100$. 
From Figure~\ref{crosOfpolarized}, one  can find that the beam polarization can significantly affect the scattering cross section of the process.  The cross section with $P_{\mu^-}=80$ is $5.2$ times larger than that with $P_{\mu^-}=-80$  when $P_{\mu^+}=-50$. The maximum cross section is about $4.5 \times 10^{-3}$ $\text{pb}$ with  $P_{\mu^+}$ close to -100 and $P_{\mu^-}$  close to 100.  Besides, there is an interesting point with $P_{\mu^+}$ around $60$, where the cross section is about $9 \times 10^{-4}~\text{pb}$ with all kinds of value for  $P_{\mu^-}$. 

\begin{figure}[tbh] 
\center
	\includegraphics[width=16cm,height=6cm]{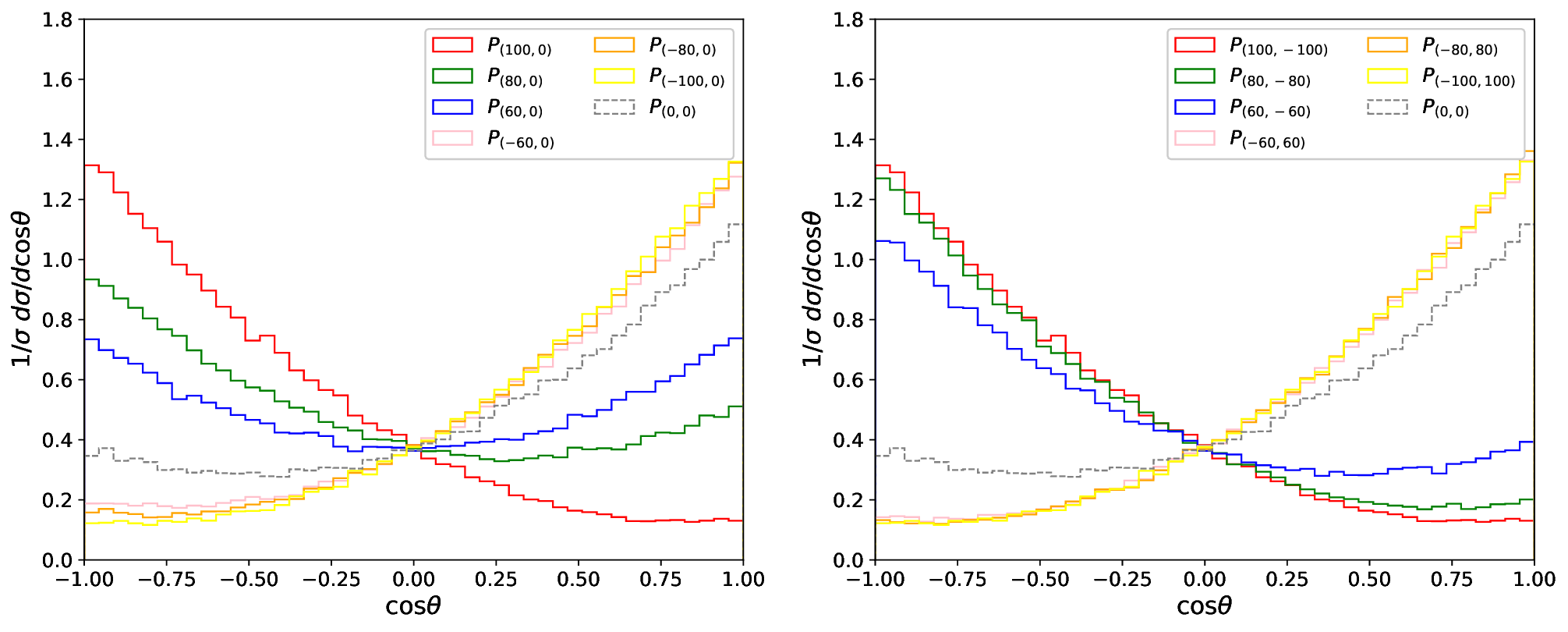}
	\caption{The angular distribution for process $\mu^+ \mu^- \rightarrow \gamma^*/Z/Z^\prime \rightarrow q \bar{q}~(q=u,c,d,s,b)$ with $Z^\prime=\sqrt{s}=6000$ GeV and $\sin\alpha = 1 \times 10^{-4}$. The colored lines correspond to various polarization values $P_{(P_{\mu^+}, P_{\mu^-})}$. }
    \label{qqPolDistribution}
\end{figure}

\begin{figure}[tbh] 
    \center
	\includegraphics[width=16cm,height=6cm]{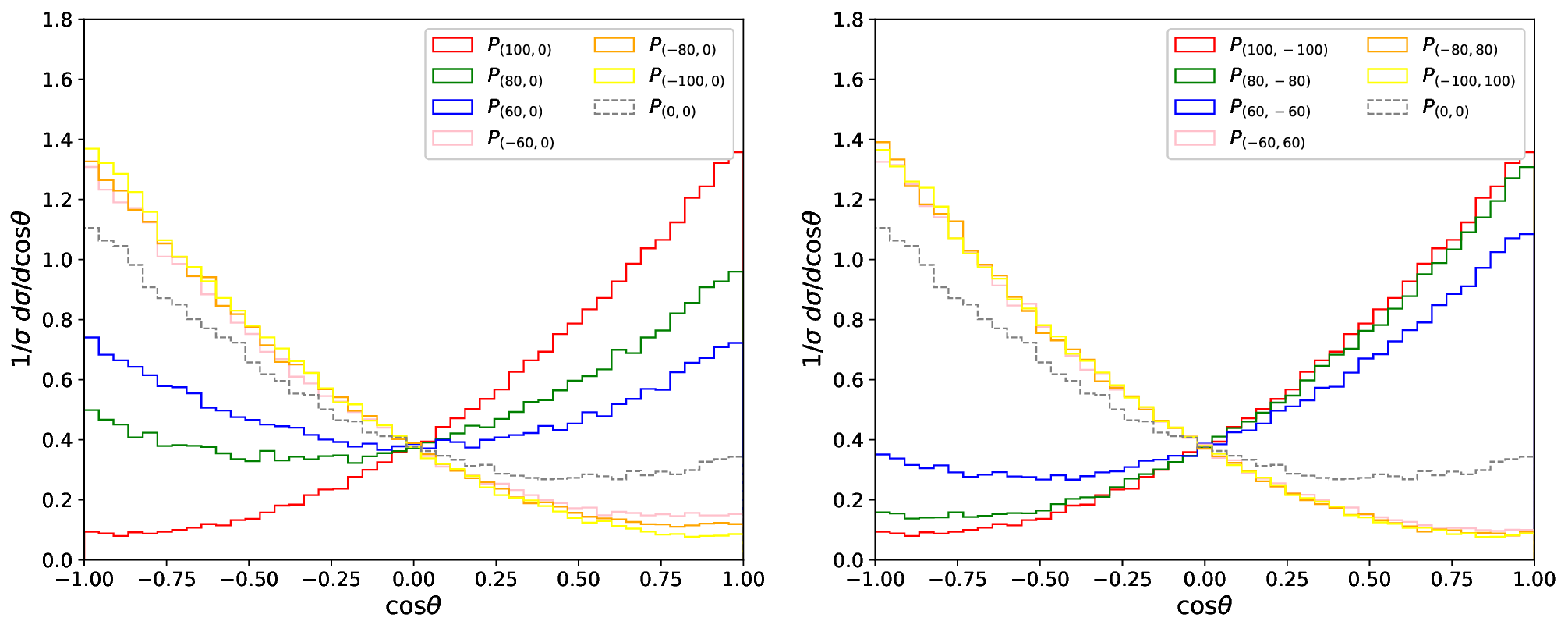}
	\caption{The angular distribution for process $\mu^+ \mu^- \rightarrow \gamma^*/Z/Z^\prime \rightarrow t \bar{t} \rightarrow b \bar{b} W^+ W^- \rightarrow b \bar{b} l^+ l^- \nu_l \bar{\nu}_l$ with $Z^\prime=\sqrt{s}=6000$ GeV and $\sin\alpha = 1 \times 10^{-4}$. The colored lines correspond to various polarization values $P_{(P_{\mu^+}, P_{\mu^-})}$. }
    \label{ttPolDistribution}
\end{figure}

We also show the  final quarks angular distribution for  $\mu^+ \mu^- \rightarrow \gamma^*/Z/Z^\prime \rightarrow q \bar{q}~(q=u,c,d,s,b)$ with different values of beam polarization in Figure~\ref{qqPolDistribution}. 
The angular distribution has a slight difference when $P_{{\mu}^+}<0, P_{{\mu}^-}=0$ comparing with the unpolarization condition. While, it can greatly affect the angular distribution when $P_{{\mu}^+}$ is positive, and even changes to the inverse tendency of the distribution when $P_{{\mu}^+}>60$.  For example, the distribution asymmetry with $P_{{\mu}^+}=60, P_{{\mu}^-}=-60$ is larger than that with $P_{{\mu}^+}=60, P_{{\mu}^-}=0$. In Figure~\ref{ttPolDistribution}, we show the dependence of the angular distribution of the final charged leptons  and the beam  polarization in the top quark pair production. We can clearly see that the beam polarization has a great influence on the angular distribution of the final particles comparing with the unpolarized condition. The distribution is the same as  in Figure~\ref{qqPolDistribution} with a mirror symmetry.

\begin{table}[H]
\begin{center}
\setlength{\abovecaptionskip}{6pt}
\setlength{\belowcaptionskip}{0pt}
\begin{tabular}{c c c c c c c c c}
\toprule
 \multicolumn{4}{c}{$\mu^+ \mu^- \rightarrow q\bar{q}$} &\qquad \quad& \multicolumn{4}{c}{$\mu^+ \mu^- \rightarrow t \bar{t} \rightarrow b \bar{b} l^+ l^- \nu_l \bar{\nu}_l$}\\
$P_{(P_{\mu^+},P_{\mu^-})}$ & $A_{FB}$ \quad& $P_{(P_{\mu^+},P_{\mu^-})}$ & $A_{FB}$\quad & & $P_{(P_{\mu^+},P_{\mu^-})}$ & $A_{FB}$ \quad & $P_{(P_{\mu^+},P_{\mu^-})}$ & $A_{FB}$ \\
\midrule
$(100,-100)$ & $-0.610$ \quad & $(100,0)$ & $-0.610$ \quad & & $(100,-100)$ & $0.654$ \quad & $(100,0)$ & $0.654$ \\
$(80,-80)$ & $-0.552$ \quad & $(80,0)$ & $-0.210$ \quad & & $(80,-80)$ & $0.594$ \quad & $(80,0)$ & $ 0.252$ \\
$(60,-60)$ & $-0.363$ \quad & $(60,0)$ & $0.003$ \quad & & $(60,-60)$ & $0.390$ \quad & $(60,0)$ & $-0.0002$ \\
$(-60,60)$ & $0.599$ \quad & $(-60,0)$ & $0.548$ \quad & & $(-60,60)$ & $-0.641$ \quad & $(-60,0)$ & $-0.590$ \\
$(-80,80)$ & $0.613$ \quad & $(-80,0)$ & $0.518$ \quad & & $(-80,80)$ & $-0.660$ \quad & $(-80,0)$ & $-0.625$ \\
$(-100,100)$ & $0.616$ \quad & $(-100,0)$ & $0.616$ \quad & & $(-100,100)$ & $-0.661$ \quad & $(-100,0)$ & $-0.660$ \\
\bottomrule
\end{tabular}
\caption{The forward-backward asymmetry with different polarization of processes $\mu^+ \mu^- \rightarrow q \bar{q} (q=u,c,d,s,b)$ and $\mu^+ \mu^- \rightarrow t \bar{t}\rightarrow b \bar{b} W^+ W^-  \rightarrow b \bar{b} l^+ l^- \nu_l \bar{\nu}_l$ with $m_{Z^\prime}=\sqrt{s}=6000$ GeV and $\sin\alpha = 1 \times 10^{-4}$. }
\label{DisTbFBAmumuToqqDecay}
\end{center}
\end{table}

The forward-backward asymmetry with different polarization of processes $\mu^+ \mu^- \rightarrow q \bar{q} (q=u,c,d,s,b)$ and $\mu^+ \mu^- \rightarrow t \bar{t}\rightarrow b \bar{b} W^+ W^-  \rightarrow b \bar{b} l^+ l^- \nu_l \bar{\nu}_l$ are listed in Table \ref{DisTbFBAmumuToqqDecay}. The asymmetry can be $-0.61(0.62)$ with the polarization of $P_{{\mu}^+}=-100, P_{{\mu}^-}=100$ ($P_{{\mu}^+}=100, P_{{\mu}^-}=-100$) in $\mu^+ \mu^- \rightarrow q \bar{q} $ process. The asymmetry of $\mu^+ \mu^- \rightarrow t \bar{t}\rightarrow b \bar{b} W^+ W^-  \rightarrow b \bar{b} l^+ l^- \nu_l \bar{\nu}_l$ is about the same value as  $\mu^+ \mu^- \rightarrow q \bar{q} $ process but with an opposite sign. The results show that the asymmetry is an  observable sensitively to the  polarization. 
%
\subsection{$Z^\prime$ search in $\mu^+ \mu^- \rightarrow l^+ l^-$ }
\subsubsection{The cross section of process $ \mu^+  \mu^- \rightarrow l^+ l^-$}
%
\begin{figure}[tbh] 
  \center
	\includegraphics[width=16cm,height=18cm]{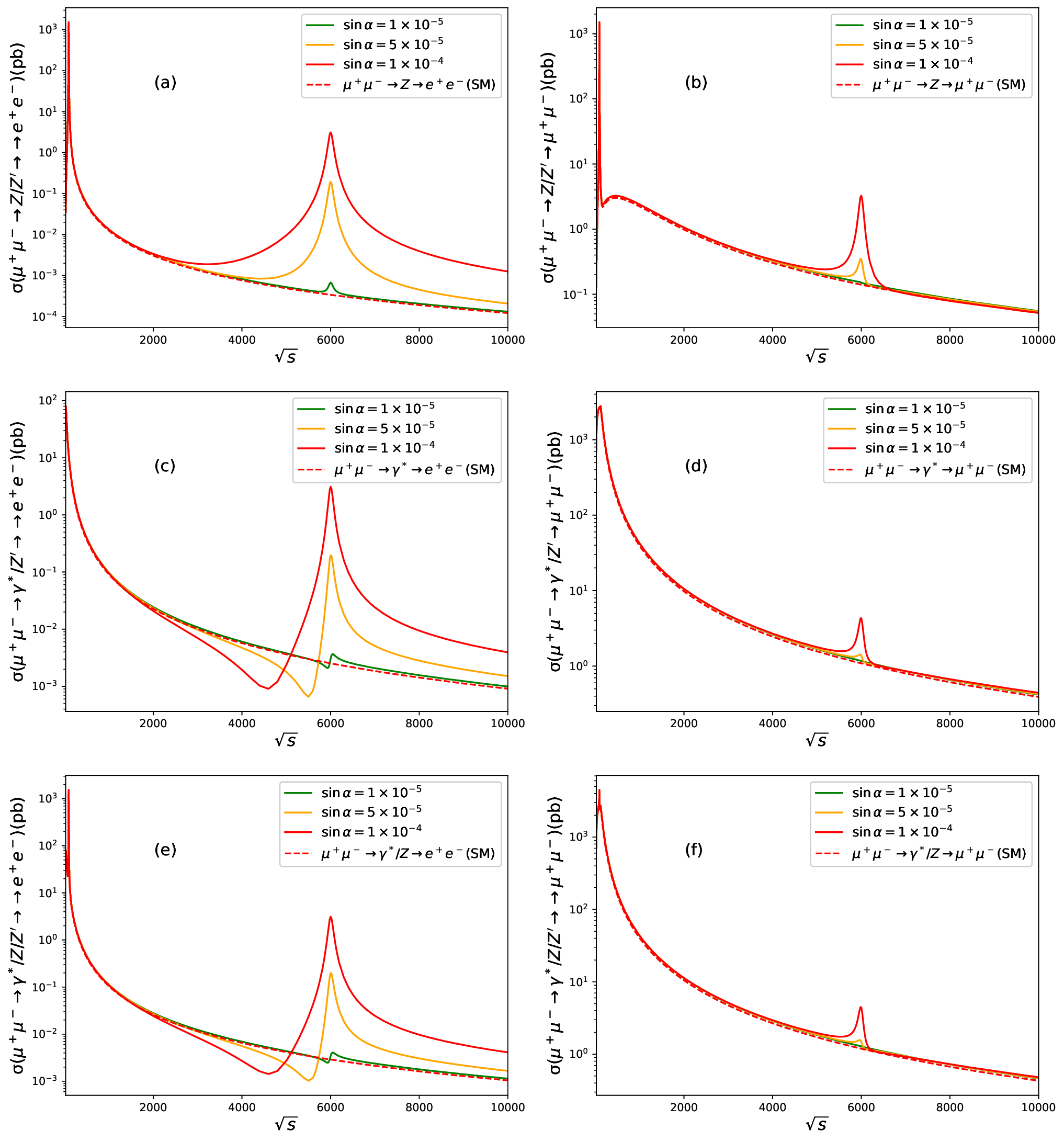} 
	\caption{The cross section of process $\mu^+ \mu^- \rightarrow \gamma^*/Z^\prime \rightarrow l^+ l^-$ with $m_{Z^{\prime}}=6000$ GeV versus $\sqrt{s}$ where the final lepton $l$ is $e$ (a), (c), (e) or $\mu$ (b), (d), (f).}
    \label{llCrossSection}
\end{figure}

As a process widely studied for the neutral gauge boson production, the process  $\mu^+ \mu^- \rightarrow l^+ l^-(l=e,\mu)$  on the future Muon Collider is also very meaningful. The dilepton process, compared with the diquark process, has additional $t-$channel Feynman diagrams. Due to the extremely short lifetime of the $\tau$ lepton, we only consider the final states with $e$ and $\mu$ referring to the dilepton final states.
For the final states with $e^+ e^-$ and $\mu^+ \mu^-$, as shown in Figure~\ref{llCrossSection}, they exhibit significant differences in the scattering cross section as well. The reason for this phenomenon is that only the $s-$channel Feynman diagram contributes to the scattering cross section in process $\mu^+ \mu^- \rightarrow e^+ e^-$, while there are also $t-$channel Feynman diagrams in process $\mu^+ \mu^- \rightarrow \mu^+ \mu^-$ that contribute to the scattering cross section in addition to the $s-$channel contribution.

The cross section distribution of process $\mu^+ \mu^- \rightarrow  e^+ e^-$ is similar to the process with diquark final state with only $s-$channel Feynman diagrams. Due to the  interference between $Z^\prime$ and $\gamma ^*$, the distribution has a valley in the region of C.M.S. energy smaller and closed to $Z^\prime$ mass pole  as shown in Figure~\ref{llCrossSection} (c), (e). However, the production of final state $\mu^+ \mu^-$ has a different distribution since the   $t-$channel Feynman diagram contributes to a positive value to offset the negative contribution from the interference between  $Z^\prime$  and $\gamma ^*$ as shown in Figure~\ref{llCrossSection} (d), (f).
\subsubsection{The angular distribution of leptons in  process $\mu^+ \mu^- \rightarrow l^+ l^-$}
\begin{figure}[tbh] 
   \center
	\includegraphics[width=16cm,height=5.5cm]{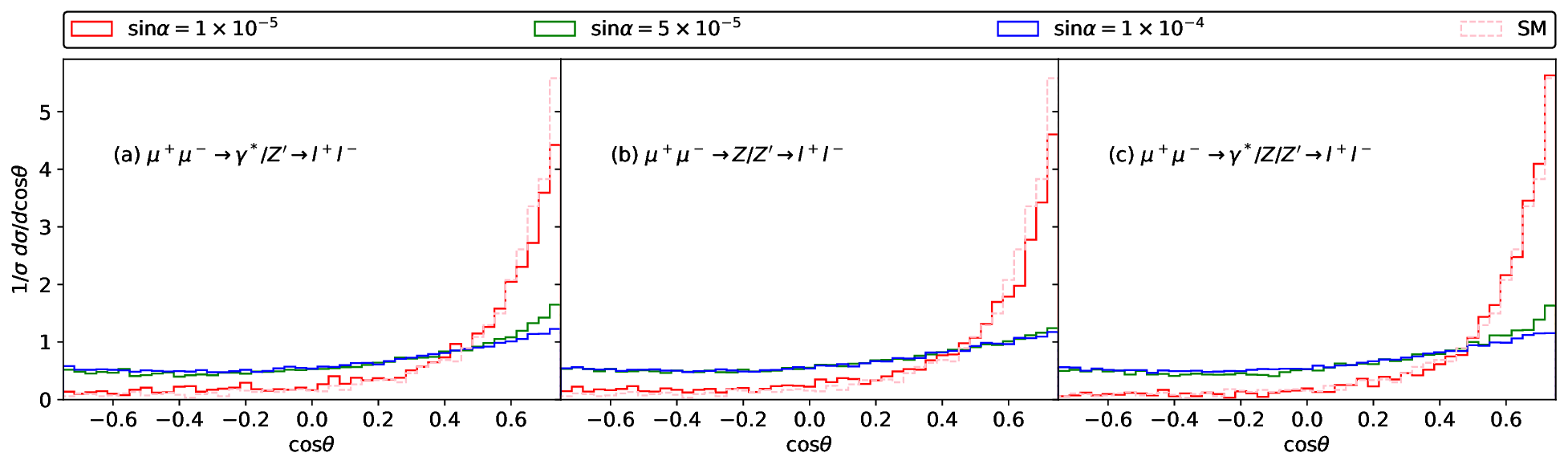}
	\caption{Angular distribution of leptons in the final state of processes $\mu^+ \mu^- \rightarrow \gamma^*/Z^\prime \rightarrow l^+ l^-$ (a), $\mu^+ \mu^- \rightarrow Z/Z^\prime \rightarrow l^+ l^-$ (b), $\mu^+ \mu^- \rightarrow \gamma^*/Z/Z^\prime \rightarrow l^+ l^-$ (c) with the C.M.S. energy and mass of $Z^\prime$ are 6000 GeV.}
    \label{llDistribution}
\end{figure}
The angular distribution of the final  leptons in  process $\mu^+ \mu^- \rightarrow l^+ l^-$ is investigated in this part. The angle $\theta$ is defined by the moving direction of $l^-$ and  initial $\mu^-$ as Equation~\eqref{eq13}. We plot the angular distribution of leptons in the final state of the dilepton process in Figure~\ref{llDistribution}. The distribution is not the same as $q\bar{q}$ because of the contribution from $\mu^+ \mu^- \rightarrow \mu^+ \mu^-$ $t-$channel contribution. Most of the negative charged final leptons go along with the direction of $\cos \theta >0$, especially large number of the  negative charged final leptons is close to the direction of $\mu^-$.  The forward-backward asymmetry is also defined as the formula of Equation~\eqref{eq14} and the numbers are listed in Table  \ref{DisFBA}. The asymmetry associated solely with the presence of $Z^\prime$ contribution exhibits a reduced magnitude in comparison to that arising solely from the SM $Z$ boson process. However, when the mixing angle is gradually increased, the contribution of $Z^\prime$ to the process increases, leading to a decrease in the overall asymmetry. The process $\mu^+ \mu^- \rightarrow l^+ l^-$ has additional contribution from $t$-channel Feynman diagram comparing with the process $\mu^+ \mu^- \rightarrow q \bar{q}$, which leads to a greater asymmetry in the  final particle angular distribution.

\begin{table}[H]
\begin{center}
\setlength{\abovecaptionskip}{6pt}
\setlength{\belowcaptionskip}{0pt}
\begin{tabular}{c c c c c c c}

\toprule
$\sin\alpha$ \qquad & $A_{FB}(Z/Z^\prime)$ \qquad & $A_{FB}(\gamma^*/Z^\prime)$ \qquad & $A_{FB}(\gamma^*/Z/Z^\prime)$ \qquad & $A_{FB}(Z)$ \qquad &  $A_{FB}(\gamma^*)$ \qquad &  $A_{FB}(\gamma^*/Z)$\\
\midrule
 $1 \times 10^{-5}$ \qquad & $0.727$ \quad & $0.746$ \qquad & $0.844$ \qquad & \multirow{3}*{0.757  \quad} & \multirow{3}*{0.732  \qquad} & \multirow{3}*{0.883  \qquad}\\

 $5 \times 10^{-5}$ \qquad & 0.240 \qquad & $0.298$ \qquad & $0.305$ \qquad & ~ & ~ & ~\\

 $1 \times 10^{-4}$ \qquad & 0.224 \qquad & $0.231$ \qquad & $0.230$ \qquad & ~ & ~ & ~\\
\bottomrule

\end{tabular}
\caption{Forward-backward asymmetry for the process $\mu^+ \mu^- \rightarrow l^+ l^-$ with $\sqrt{s}=m_{Z^\prime}=6000$ GeV.}
\label{DisFBA}
\end{center}
\end{table}
\subsubsection{Signals with beam polarization}
\begin{figure}[tbh] 
  \center
	\includegraphics[width=16cm,height=6cm]{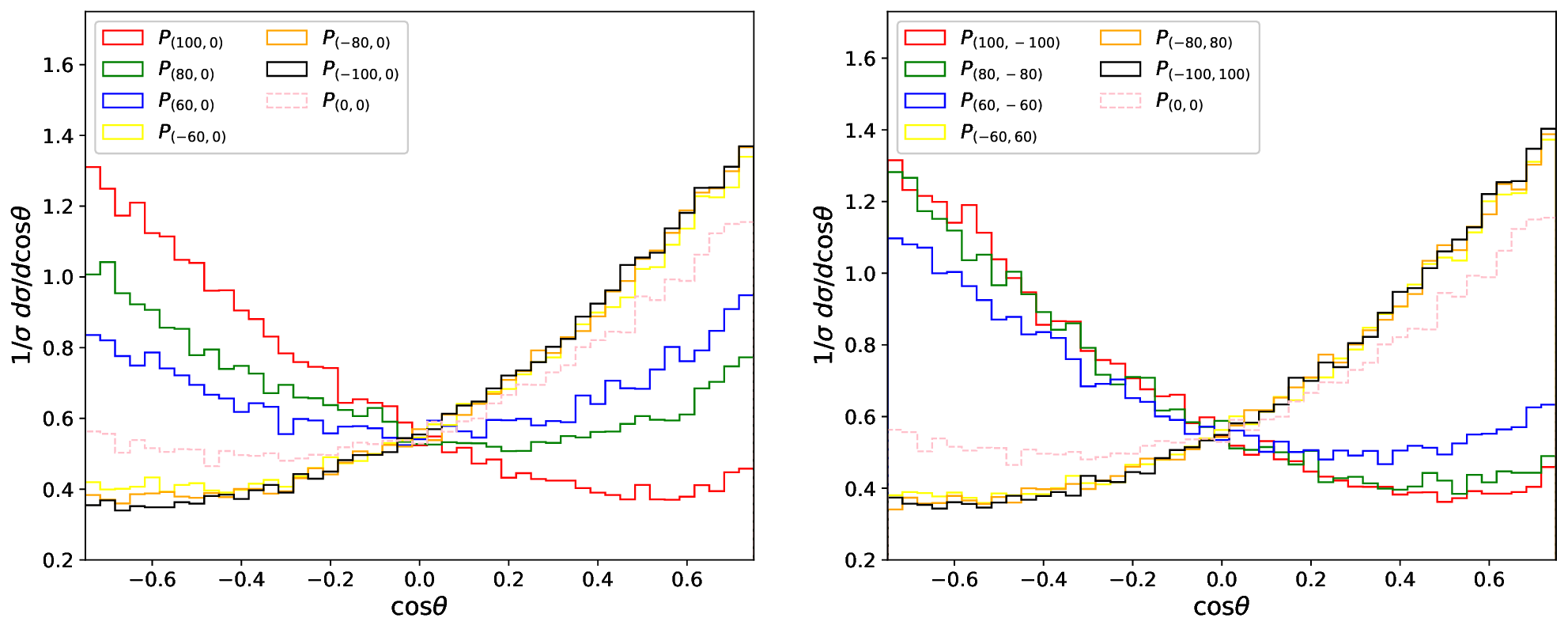}
	\caption{The angular distribution for process $\mu^+ \mu^- \rightarrow \gamma^*/Z/Z^\prime \rightarrow l^+ l^-$ with $Z^\prime=\sqrt{s}=6000$ GeV and $\sin\alpha = 1 \times 10^{-4}$. The colored lines correspond to various polarization values $P_{(P_{\mu^+}, P_{\mu^-})}$. }
    \label{llPol}
\end{figure}

 We investigate  the  relationship between the  angular distribution of final lepton and the polarization of the initial muons. Figure~\ref{llPol} shows that the various beam polarizations will indeed change the angular distribution of final leptons.  The tendency of the angular distribution with $P_{\mu^+}=100,P_{\mu^-}=0$ is inverse with $P_{\mu^+}=-100,P_{\mu^-}=0$.  In Table \ref{tb6}, we list the values of forward-backward asymmetry with different beam polarization. The asymmetry is defined as the formula of Equation~\eqref{eq14}.  It is found that when the value of $P_{\mu^+}$ is negative in the dilepton process, the forward-backward asymmetry for the final lepton angular distribution will change slightly. The asymmetry can be -0.361 with $P_{\mu^+}=100,P_{\mu^-}=-100$ and 0.386 with $P_{\mu^+}=-100,P_{\mu^-}=100$ . These results show that the study of the $Z'$ production with beam polarization is worthy to investigate the new physics models.
\begin{table}[H]
\begin{center}
\setlength{\abovecaptionskip}{6pt}
\setlength{\belowcaptionskip}{0pt}
\begin{tabular}{ c c c c c c c c}

\toprule
 $P_{(P_{\mu^+},P_{\mu^-})}$ & $A_{FB}$ \quad & $P_{(P_{\mu^+},P_{\mu^-})}$ & $A_{FB}$ \quad  & $P_{(P_{\mu^+},P_{\mu^-})}$ & $A_{FB}$ \quad  & $P_{(P_{\mu^+},P_{\mu^-})}$ & $A_{FB}$ \\
\midrule
$(100,-100)$ & $-0.361$ \quad  & $(100,0)$ & $-0.349$ \quad  & $(-100,100)$ & $0.386$ \quad  & $(-100,0)$ & $0.380$\\
$(80,-80)$ & $-0.326$ \quad  & $(80,0)$ & $-0.126$ \quad  & $(-80,80)$ & $0.377$ \quad  & $(-80,0)$ & $0.365$\\
$(60,-60)$ & $-0.206$ \quad  & $(60,0)$ & $0.019$ \quad  & $(-60,60)$ & $0.367$ \quad  & $(-60,0)$ & $0.340$\\
\bottomrule

\end{tabular}
\caption{The forward-backward asymmetry with different polarization for process $\mu^+ \mu^- \rightarrow l^+ l^-$ with $m_{Z^\prime}=\sqrt{s}=6000$ GeV and $\sin\alpha=1 \times 10^{-4}$.}
\label{tb6}
\end{center}
\end{table}

\subsection{$Z^\prime$ search in $\mu^+ \mu^- \rightarrow Z H$ }
The study of Higgs boson properties will be an important aim at the future muon colliders. Although the branching ratio of $Z^\prime \to ZH$ is a few percent, it is meaningful to study the process of   $\mu^+ \mu^- \rightarrow Z/ Z^\prime \to Z H$. 
\subsubsection{The cross section of process $ \mu^+  \mu^- \rightarrow ZH$}
\begin{figure}[tbh] 
  \includegraphics[width=16cm,height=6cm]{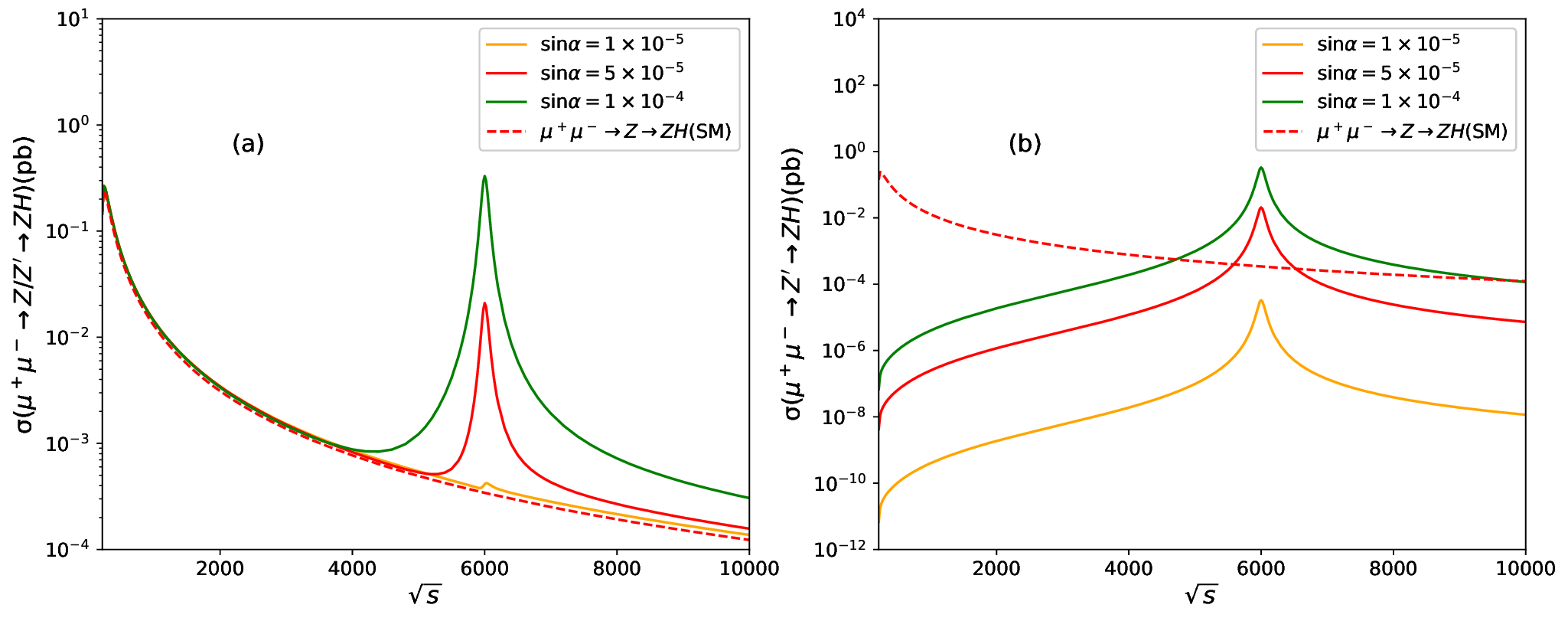}
	\caption{The cross section of process $\mu^+ \mu^- \rightarrow Z/Z^\prime \rightarrow ZH$(a) and $\mu^+ \mu^- \rightarrow Z^\prime \rightarrow ZH$(b) versus $\sqrt{s}$ with $m_{Z^{\prime}}=6000$ GeV.}
    \label{ZHCrossSection}
\end{figure}

The cross section of  $\mu^+ \mu^- \rightarrow Z^\prime \rightarrow ZH$ with various  C.M.S. energy is  shown in  Figure~\ref{ZHCrossSection}. The interference effect with $Z$ boson is included in  Figure~\ref{ZHCrossSection} (a), which shows the same distributions as in electron-positron collider~\cite{Yin:2021rlr}.   
One can notice  that when the C.M.S. energy is close to the mass of $Z^\prime$, a large resonance peak will be generated and the size of the resonance peak is affected by the mixing angle $\sin\alpha$. When $\sin\alpha>1\times 10^{-5}$, the peak can be observed significantly. But the contribution of $Z^\prime$ to the total cross section is small with the collision energy far away from $Z^\prime$ mass pole. This phenomenon implies the signal of new physics is outstanding with the collision energy close to the mass of $Z^\prime$, which shows the advantage of muon collider to detect new physics with a larger mass scale than the electron-positron collider.

\subsubsection{The angular distribution of leptons in the final state of process $\mu^+ \mu^- \rightarrow ZH \rightarrow l^+ l^- b \bar{b}$}
\begin{figure}[tbh] 
 \centering
\includegraphics[width=5cm,height=3.2cm]
{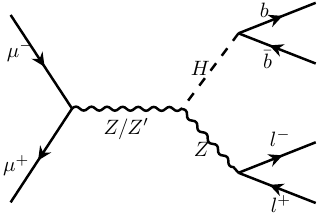}
\caption{Feynman diagram of process $\mu^+ \mu^- \rightarrow ZH \rightarrow l^+ l^- b \bar{b}$}
\label{ZHDecayFeyDia}
\end{figure}
In order to study the property of $Z^\prime Z H$ interaction, we consider the further decay of $Z$  and Higgs bosons. The angular distribution of final leptons is studied in process of  $\mu^+ \mu^- \rightarrow ZH \rightarrow l^+ l^- b \bar{b}$ with $H\rightarrow  b \bar{b}$ and $ Z \rightarrow l^+ l^- $ ($l =e, \mu$). The corresponding Feynman diagram is shown in Figure~\ref{ZHDecayFeyDia}.  
\begin{figure}[tbh] 
\center
	\includegraphics[width=8cm,height=6cm]{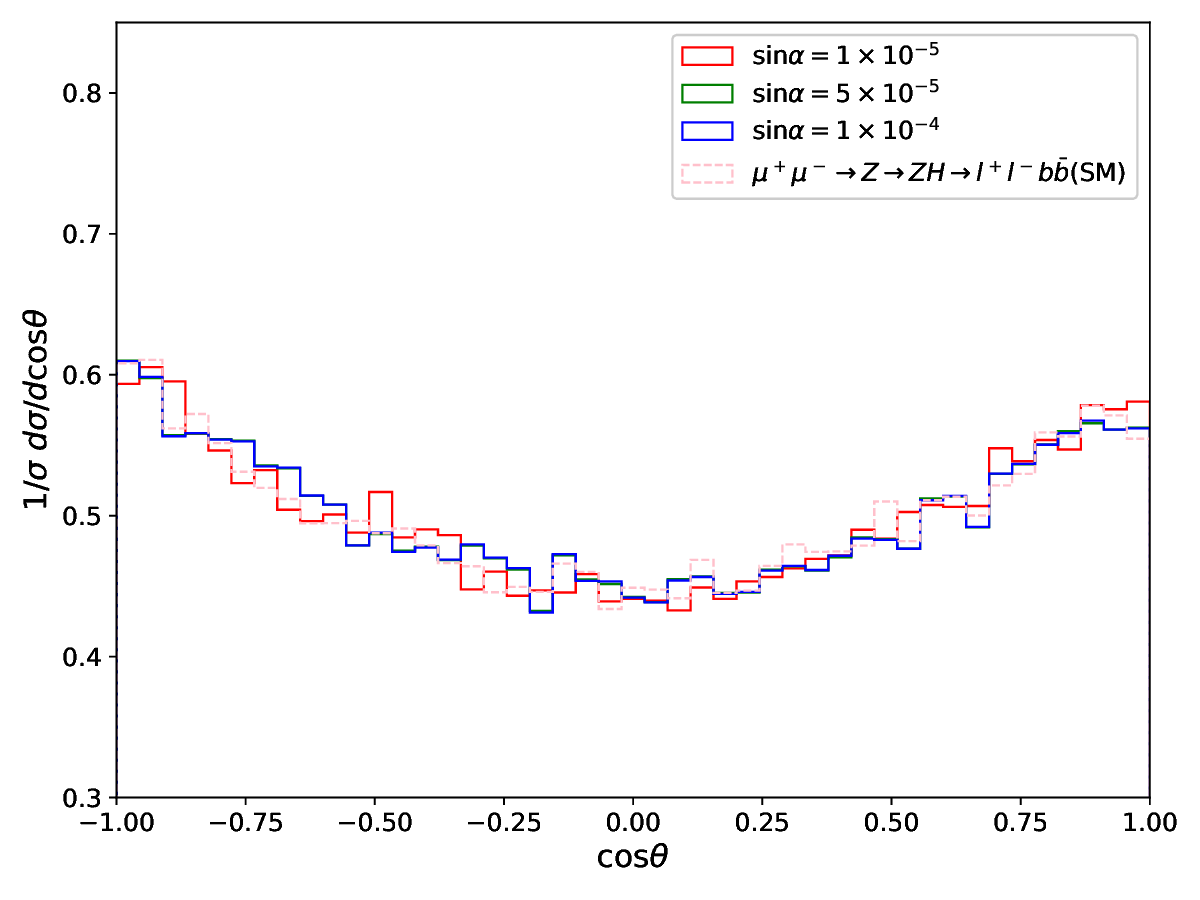}
	\caption{Angular distribution of leptons in the final state of processes $\mu^+ \mu^- \rightarrow Z/Z^\prime \rightarrow ZH \rightarrow l^+ l^- b \bar{b}$ with the C.M.S. energy and mass of $Z^\prime$ are 6000 GeV.}
    \label{ZHDistribution}
\end{figure}
We present the angular distribution of the final leptons for $\mu^+ \mu^- \rightarrow ZH \rightarrow l^+ l^- b \bar{b}$ process in  Figure~\ref{ZHDistribution} . It is worthy of note that in order to better observe the angular distribution of the final leptons, we change the center of mass system of leptons into the center of mass system of $Z$ through Lorentz transformation when calculating the scattering angle as~\cite{Yin:2021rlr}. In the calculation, we still use Formula \eqref{eq13} to calculate the angular distribution of the final leptons, where $\bm{p_{i}}$ is $\bm{p_{\mu^-}}$ and $\bm{p_{f}}^*$ is $\bm{p_{l^-}}^*$. The asymmetry of this process is relatively small. When $\sin\alpha$ is $1 \times 10^{-5}$, $5 \times 10^{-5}$ and $1 \times 10^{-4}$, the forward-backward asymmetries are $-0.027$, $-0.0092$ and $-0.0094$ respectively with  $Z^\prime$ contributions while it is $-0.022$ for the SM process.  Although the difference of the angular distribution is not as  clear as those in reference~\cite{Yin:2021rlr}, the investigated mass range of $Z^\prime $ reaches to a few TeV and  $\sin\alpha$ is smaller about two order of magnitudes compared to the previous study. We will study the effect of beam polarization on this process in the following section.
\subsubsection{Signals with beam polarization}
\begin{figure}[tbh] 
\center
	\includegraphics[width=16cm,height=6cm]{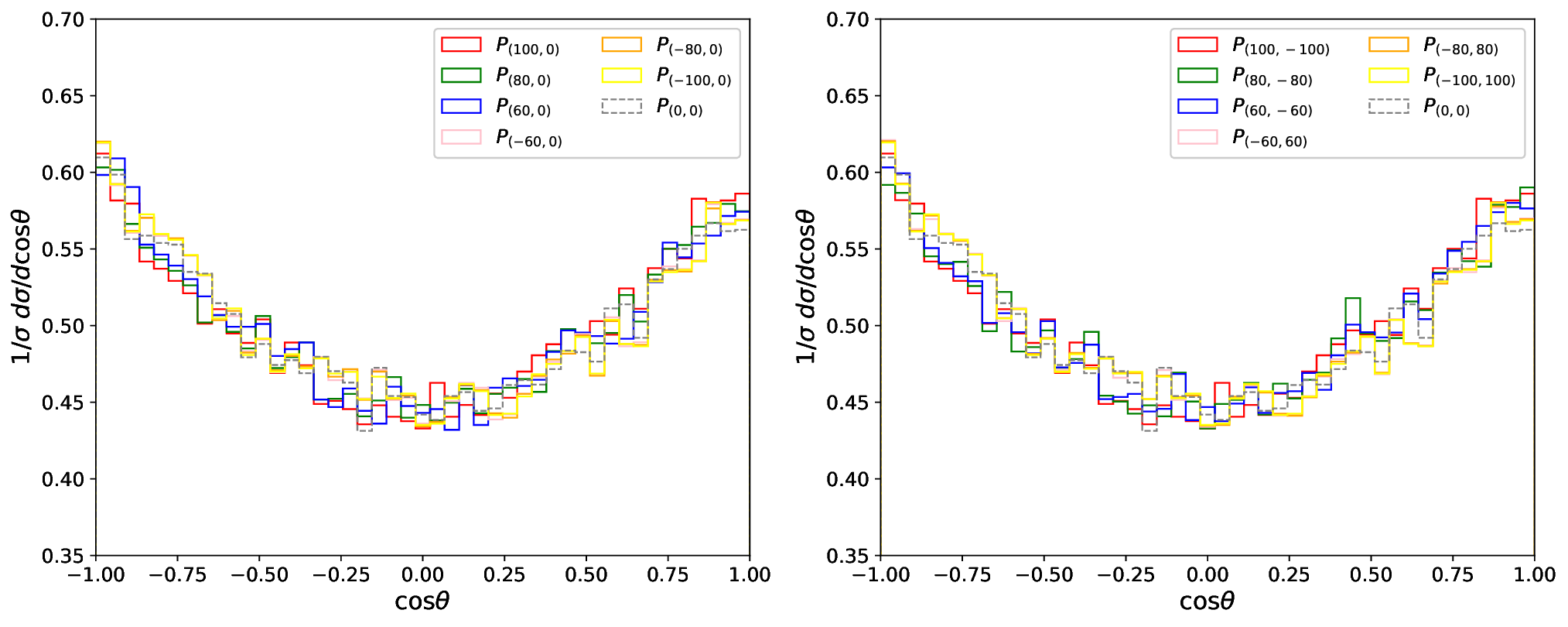}
	\caption{The angular distribution for process $\mu^+ \mu^- \rightarrow Z/Z^\prime \rightarrow ZH \rightarrow  l^+ l^- b \bar{b}$ with $Z^\prime=\sqrt{s}=6000$ GeV and $\sin\alpha = 1 \times 10^{-4}$. The colored lines correspond to various polarization values $P_{(P_{\mu^+}, P_{\mu^-})}$.}
    \label{ZHPOlDistribution}
\end{figure}
\begin{table}[H]
\begin{center}
\setlength{\abovecaptionskip}{6pt}
\setlength{\belowcaptionskip}{0pt}
\begin{tabular}{c c c c c c c c}
\toprule
 $P_{(P_{\mu^+},P_{\mu^-})}$  & $A_{FB}$ \qquad & $P_{(P_{\mu^+},P_{\mu^-})}$ & $A_{FB}$ \qquad & $P_{(P_{\mu^+},P_{\mu^-})}$ & $A_{FB}$ \qquad & $P_{(P_{\mu^+},P_{\mu^-})}$ & $A_{FB}$ \\
\midrule
$(100,-100)$ & $0.009$ \qquad & $(100,0)$ & $0.009$ \qquad & $(-100,100)$ & $-0.014$ \qquad & $(-100,0)$ & $-0.014$\\
$(80,-80)$ & $0.004$ \qquad & $(80,0)$ & $0.001$ \qquad & $(-80,80)$ & $-0.014$ \qquad & $(-80,0)$ & $-0.015$\\
$(60,-60)$ & $0.002$ \qquad & $(60,0)$ & $-0.005$ \qquad & $(-60,60)$ & $-0.015$ \qquad & $(-60,0)$ & $-0.013$\\
\bottomrule
\end{tabular}
\caption{The forward-backward asymmetry with various polarizations of process $\mu^+ \mu^- \rightarrow ZH \rightarrow l^+ l^- b \bar{b}$ with $m_{Z^\prime}=\sqrt{s}=6000$ GeV and $\sin\alpha=1 \times 10^{-4}$.}
\label{tb7}
\end{center}
\end{table}
The influence of polarization on the  cross section of $\mu^+ \mu^- \rightarrow Z/Z^\prime \rightarrow ZH \rightarrow  l^+ l^- b \bar{b}$ is studied. 
The final lepton angular distributions with different beam polarizations are shown in Figure~\ref{ZHPOlDistribution}. The corresponding  forward-backward asymmetries are listed in Table \ref{tb7}. The  asymmetry can change sign with some specific polarizations, and the absolute values of the asymmetry  range within a few percent. It requires a high precision detection on the asymmetry to distinguish the new physics through this process in the future muon collider.  
\subsection{$Z^\prime$ search in $\mu^+ \mu^- \rightarrow W^+ W^-$ }
The interaction between $Z^\prime$ and charged gauge boson is also an important part for the study of $Z^\prime$ boson. In this section, we investigate the properties of the new physics model in the process of $\mu^+ \mu^- \rightarrow W^+ W^-$ at the future muon collider.  
\subsubsection{The cross section of process $ \mu^+  \mu^- \rightarrow W^+ W^-$}
\begin{figure}[tbh] 
\centering
	\subfigure[]{\includegraphics[width=4.5cm,height=3.2cm]
{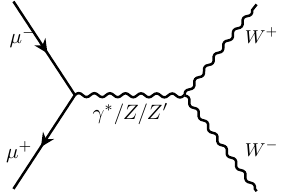}}
\hspace{0.5in}
\subfigure[]{\includegraphics[width=4cm,height=3.2cm]
{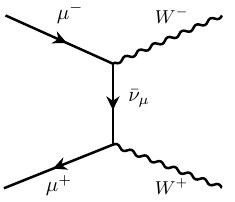}}
	\caption{Feynman diagram of $\mu^+ \mu^- \rightarrow W^+ W^-$ with $s-$channel (a) and $t-$channel (b).}
    \label{WWFeymanDia}
\end{figure}

The Feynman diagrams of $ \mu^+  \mu^- \rightarrow W^+ W^-$ are shown in Figure~\ref{WWFeymanDia}.  Each diagram separately contributes to a cross section that grows in the same manner as
\begin{equation} \label{eq16}
\frac{\text{d}\sigma}{\text{dcos}\theta}(\mu^+\mu^-\rightarrow W^+W^-)\sim \frac{\pi \alpha^2}{4s}\cdot \left|\varepsilon(k_+)\cdot \varepsilon(k_-) \right|^2,
\end{equation}
where $k_+$ and $k_-$ are the momenta of the outgoing $W$ bosons.
The contribution to the cross section within only $s-$channel Feynman diagram has a bad behavior at large C.M.S. energy which will break the unitarity. Fortunately, the badly behaved terms might be cancelled including  the $t-$channel diagrams, leaving a more proper high-energy behavior. The cancellation is required by the Ward identity of the gauge theory. We calculate the cross section of $ \mu^+  \mu^- \rightarrow W^+ W^-$ process using FeynArts and FeynClac codes~\cite{Hahn:2000kx,Shtabovenko:2016sxi,Shtabovenko:2020gxv}. The detailed calculations are given in Appendix B.
\begin{figure}[tbh] 
\centering
\includegraphics[width=8cm,height=6cm]{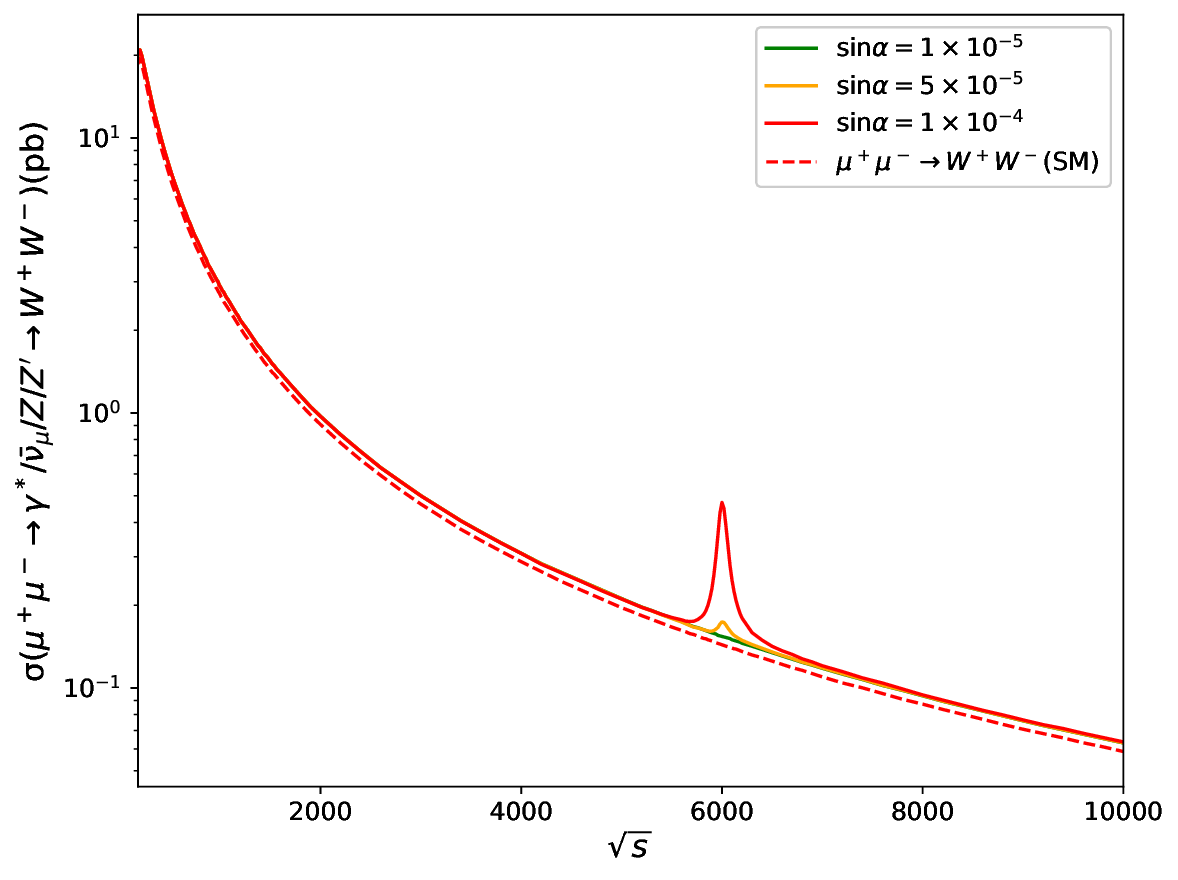}
\caption{The cross section of $\mu^+ \mu^- \rightarrow \gamma^*/\bar{\nu}_\mu/Z/Z^\prime \rightarrow W^+ W^-$ versus $\sqrt{s}$ with $M_{Z^\prime}=6000$ GeV.}
\label{WWCro}
\end{figure}
The cross section of  $ \mu^+  \mu^- \rightarrow W^+ W^-$ with the C.M.S. energy is displayed in Figure~\ref{WWCro}. The cross section can reach $0.47$ $\text{pb}$ at the resonance peak of 6000 GeV with $\sin\alpha =1\times 10^{-4}$. The $s-$channel process and $t-$channel process are not displayed separately as the previous processes since it is unphysical with the unitarity broken.  The resonance peak is obviously compared to the purely SM contribution with $\sin\alpha > 5\times 10^{-5}$ in Figure~\ref{WWCro}.
\subsubsection{The angular distribution of leptons in $\mu^+ \mu^- \rightarrow W^+ W^- \rightarrow l^+ l^- \nu_l \bar{\nu}_l$}
\begin{figure}[tbh] 
\centering
	\subfigure[]{\includegraphics[width=4.5cm,height=3cm]{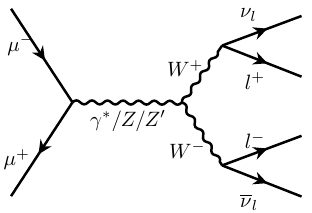}}
 \hspace{0.5in}
     \subfigure[]{\includegraphics[width=4cm,height=3.2cm]{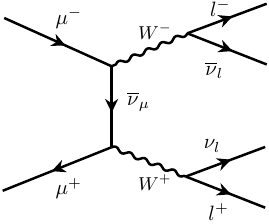}}
	\caption{Feynman diagram of $\mu^+ \mu^- \rightarrow W^+ W^- \rightarrow l^+ l^- \nu_l \bar{\nu}_l$ with $s$-channel (a) and $t$-channel (b).}
    \label{WWDeFeymanDia}
\end{figure}
\begin{figure}[tbh] 
\centering
\includegraphics[width=8cm,height=6cm]{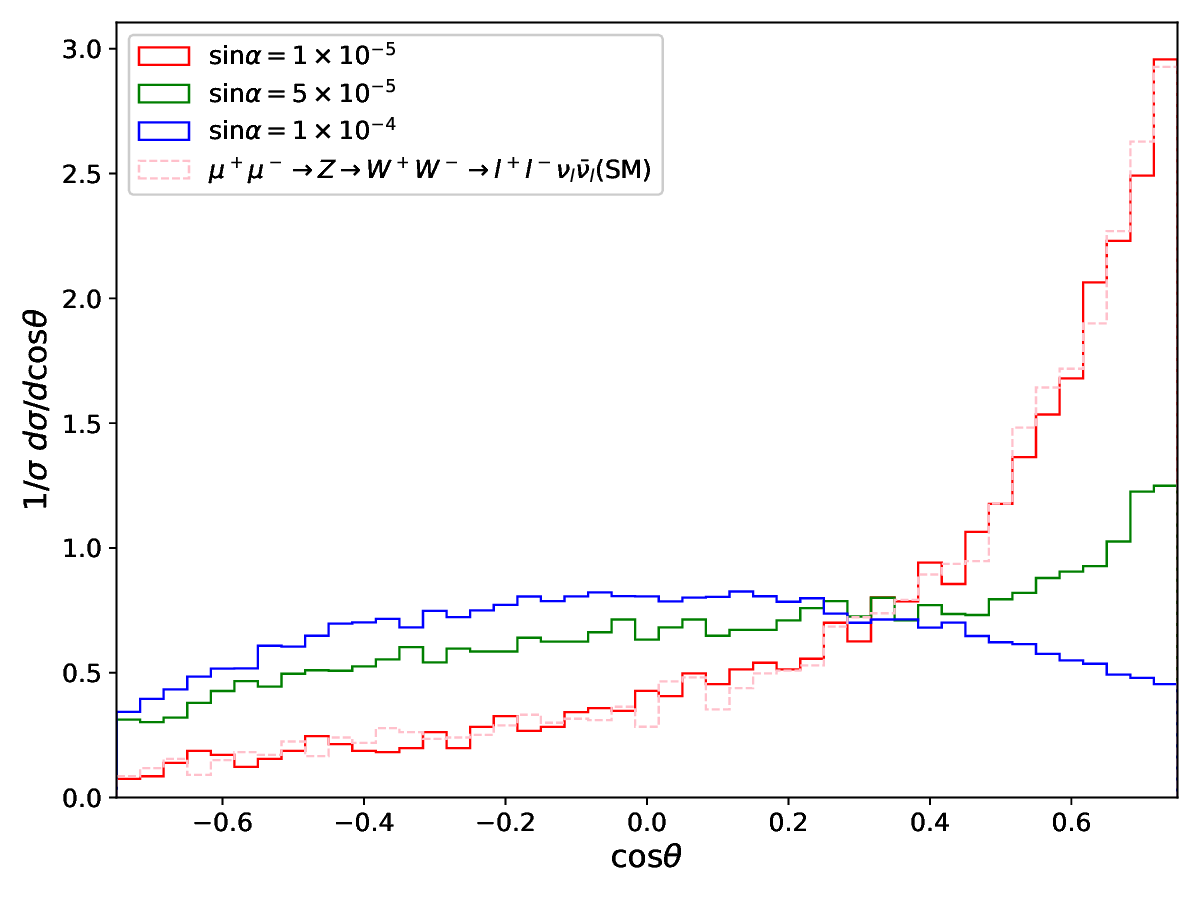}
\caption{The angular distribution for process $\mu^+ \mu^- \rightarrow \gamma^*/\bar{\nu}_\mu/Z/Z^\prime \rightarrow W^+ W^-\rightarrow l^+ l^- \nu_l \bar{\nu}_l$ with $\sqrt{s}=M_{Z^\prime}=6000$ GeV.}
\label{WWDis}
\end{figure}
 We  study the final lepton angular distribution of $\mu^+ \mu^- \rightarrow W^+ W^- \rightarrow l^+ l^- \nu_l \bar{\nu}_l$ ($l=e, \mu$) process. The Feynman diagram of this process is shown in Figure~\ref{WWDeFeymanDia}. The scattering angle is defined by the final state  $l^-$ and initial beam $\mu^-$ moving direction in the laboratory frame as formula \eqref{eq13}, where $\bm{p_{f}^*}=\bm{p_{l^-}^*} $ and $\bm{p_{i}}=\bm{p_{\mu^-}}$. We show the corresponding angular distribution of this process in Figure~\ref{WWDis}. The plots show that the distribution including $Z^{\prime}$ contribution  is the same as that of the pure SM with $\sin\alpha =1 \times 10^{-5}$, while the distribution changes greatly with the increasing of  $\sin\alpha$.  According to formula \eqref{eq14} with $|\cos\theta|<0.75 $, the forward-backward asymmetries are 0.664, 0.219, 0.015  corresponding to $\sin \alpha=1\times 10^{-5}$, $\sin \alpha=5\times 10^{-5}$, $\sin \alpha=1\times 10^{-4}$  respectively and the value from SM is 0.657.
\subsubsection{Signals with beam polarization}
\begin{figure}[tbh] 
\centering
\includegraphics[width=8cm,height=6cm]{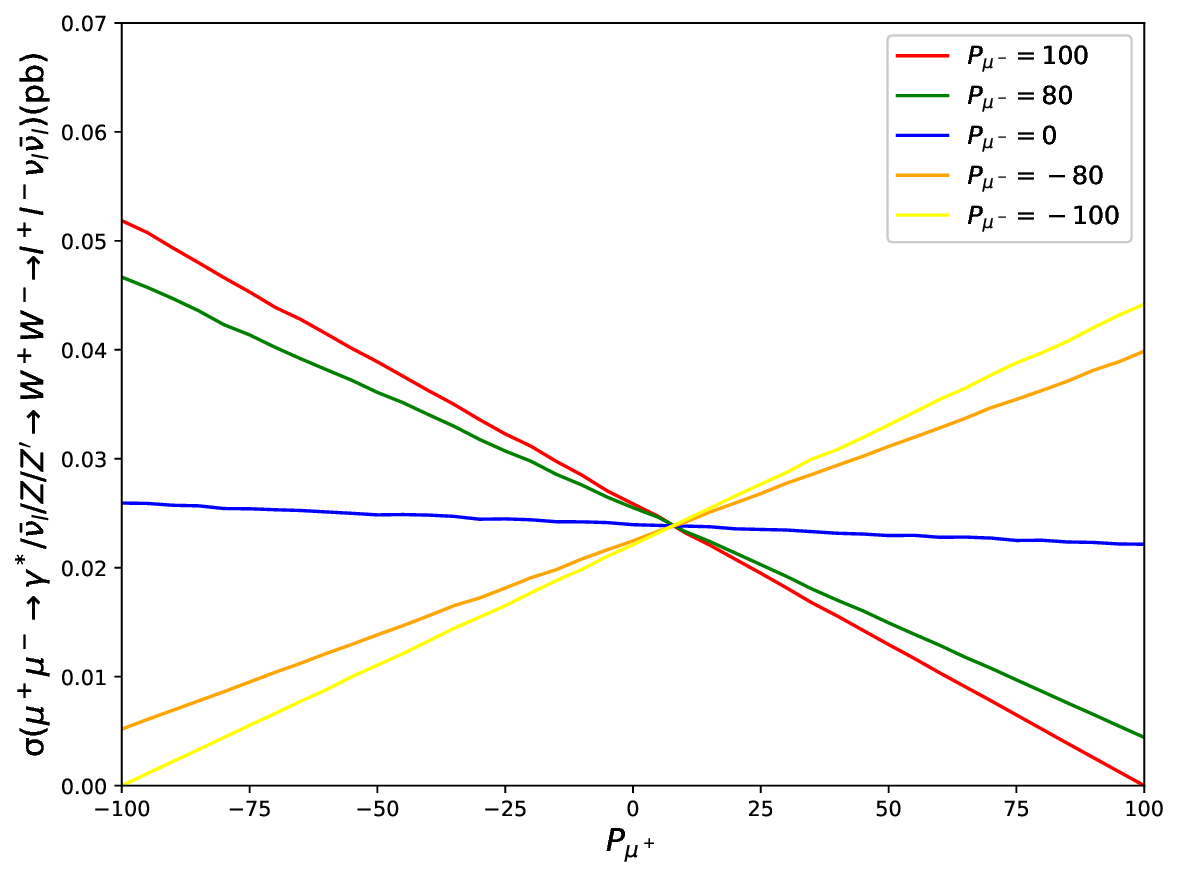}
\caption{Cross section of $\mu^+ \mu^- \rightarrow \gamma^*/\bar{\nu}_\mu/Z/Z^\prime \rightarrow W^+ W^-\rightarrow l^+ l^- \nu_l \bar{\nu}_l$ with polarized muon beam.}
\label{WWCroPol}
\end{figure}
Beam polarization has a great influence on process $\mu^+ \mu^- \rightarrow  W^+ W^-\rightarrow l^+ l^- \nu_l \bar{\nu}_l$. In the following, we will show the distribution of cross section and the final lepton angular distribution with different polarizations. 
The effect of polarization on the scattering cross section of process $\mu^+ \mu^- \rightarrow W^+ W^-$ is different from that of process $\mu^+ \mu^- \rightarrow q \bar{q}/l^+l^-$  and $\mu^+ \mu^- \rightarrow ZH$, as shown in Figure~\ref{WWCroPol}. The cross section can reach the maximum value 0.044 pb with the polarization of $P_{\mu^-}=100$, $P_{\mu^+}=-100$.  Inversely, it leads to the minimum value of the cross section  when $P_{\mu^-}=-100$, $P_{\mu^+}=-100$ and  $P_{\mu^-}=100$, $P_{\mu^+}=100$.
\begin{figure}[tbh] 
\center
	\includegraphics[width=16cm,height=6cm]{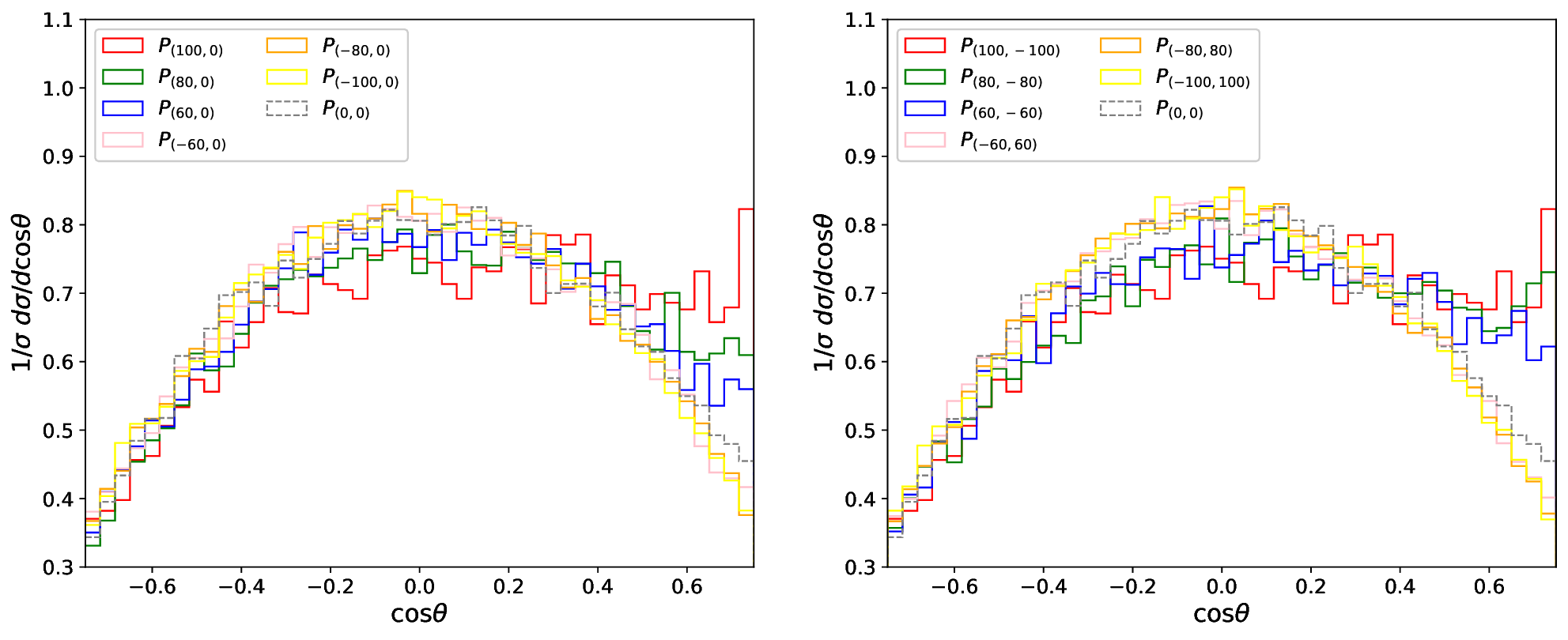}
	\caption{The angular distribution for process $\mu^+ \mu^- \rightarrow \gamma^*/\bar{\nu}_{\mu}/Z/Z^\prime \rightarrow W^+ W^- \rightarrow  l^+ l^- \nu_{l} \bar{\nu}_{l}$ with $Z^\prime=\sqrt{s}=6000$ GeV and $\sin\alpha = 1 \times 10^{-4}$. The colored lines correspond to various polarization values $P_{(P_{\mu^+}, P_{\mu^-})}$.}
    \label{WWPOlDistribution}
\end{figure}
The angular distribution of process $\mu^+ \mu^- \rightarrow  W^+ W^-\rightarrow l^+ l^- \nu_l \bar{\nu}_l$ with different polarizations are shown in Figure~\ref{WWPOlDistribution}. The corresponding forward-backward asymmetries with $|\cos\theta|<0.75$ are given in Table \ref{tb8}. The results show that the angular distributions have large discrepancy in the region of $\cos\theta>0.4$ with different beam polarization. There will be a large forward-backward asymmetry when $P_{\mu^+}>0$, while it will be significantly reduced with  $P_{\mu^+}<0$.  The forward-backward asymmetry can reach 0.086 when $P_{\mu^+}=100$, $P_{\mu^-}=-100$.

\begin{table}[H]
\begin{center}
\setlength{\abovecaptionskip}{6pt}
\setlength{\belowcaptionskip}{0pt}
\begin{tabular}{c c c c c c c c}
\bottomrule
 $P_{(P_{\mu^+},P_{\mu^-})}$  & $A_{FB}$ \quad & $P_{(P_{\mu^+},P_{\mu^-})}$ & $A_{FB}$ \quad & $P_{(P_{\mu^+},P_{\mu^-})}$ & $A_{FB}$ \quad & $P_{(P_{\mu^+},P_{\mu^-})}$ & $A_{FB}$ \\
\midrule
$(100,-100)$ & $0.086$ \quad & $(100,0)$ & $0.086$ \quad & $(-100,100)$ & $-0.001$ \quad & $(-100,0)$ & $-0.001$\\
$(80,-80)$ & $0.079$ \quad & $(80,0)$ & $0.062$ \quad & $(-80,80)$ & $-0.001$ \quad & $(-80,0)$ & $0.003$\\
$(60,-60)$ & $0.056$ \quad & $(60,0)$ & $0.038$ \quad & $(-60,60)$ & $-0.004$ \quad & $(-60,0)$ & $0.002$\\
\bottomrule
\end{tabular}
\caption{The forward-backward asymmetry with various polarizations of process $\mu^+ \mu^- \rightarrow W^+ W^- \rightarrow l^+ l^- \nu_l \bar
{\nu}_l$ with $m_{Z^\prime}=\sqrt{s}=6000$ GeV and $\sin\alpha=1 \times 10^{-4}$.}
\label{tb8}
\end{center}
\end{table}

\section{Sensitivity}

\begin{figure}[tbh] 
\center
	\includegraphics[width=16cm,height=5.3cm]{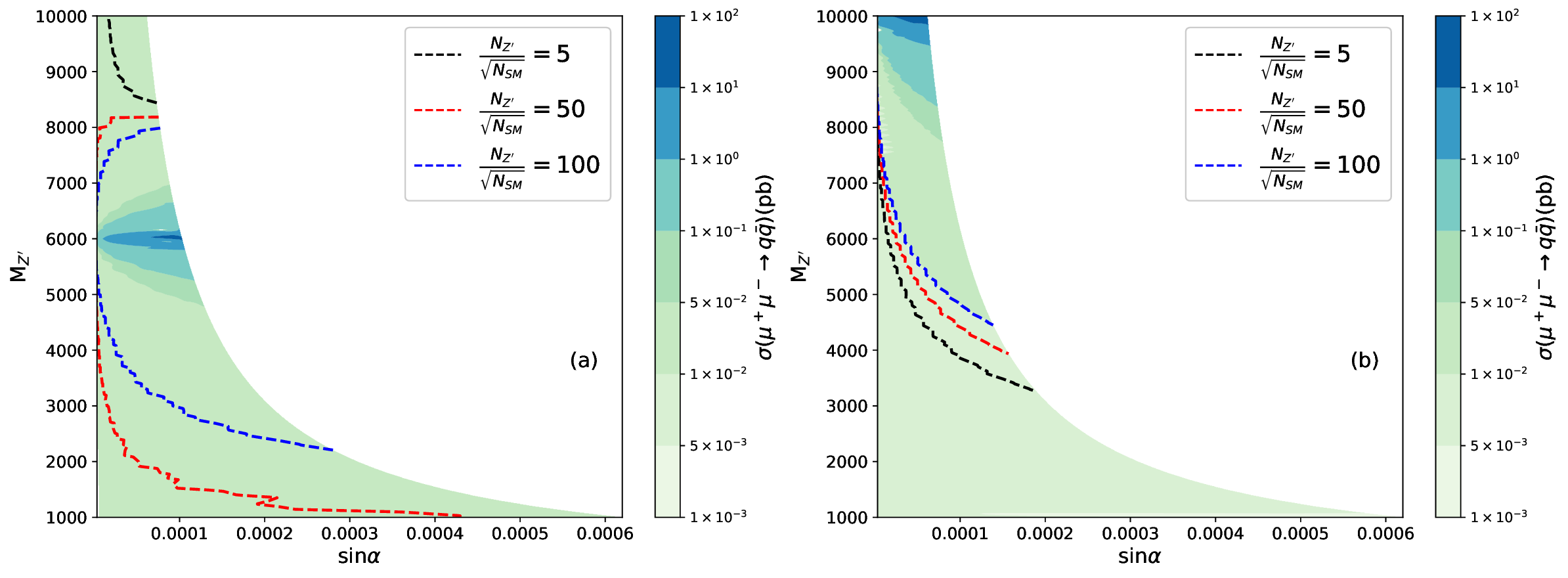}
	\caption{The contour plots of $\sin\alpha$ and $M_{Z^\prime}$ at $\sqrt{s}=6~\text{TeV} $ (a) and $10~\text{TeV} $ (b). It shows the cross section of process $\mu^+ \mu^- \to q \bar{q}$ in the corresponding parameter space. Colored dashed lines represent significance levels of 5, 10, and 100, respectively.}
    \label{contourLim}
\end{figure}

After the detailed studies of the cross section, decay channel and polarization effects, we perform the sensitivity of the $Z^\prime$ search at the muon collider.  From Figure \ref{qqdifAlphaZp} and Figure \ref{CrossMuMuToqq} we can find that the significance (defined as $N_{Z^\prime}/ \sqrt{N_{SM}}$) is large when the resonance mass is close to the collision energy. We have imposed constraints on processes in different parameter spaces in HAHM. 
In Figure \ref{contourLim},  we give the cross section distributions for  process $\mu^+ \mu^- \to q \bar{q}$ correspond to different mixing angles and $Z^\prime$ masses at the collision energy of 6 TeV and 10 TeV . The colored region imposes the constraints from $Z$ boson mass measurement. The significance lines are  illustrated with 5, 10, and 100. It can be observed that when the C.M.S energy closes to the mass of $Z^\prime$, the scattering cross section significantly increases. Furthermore, the cross section gradually decreases symmetrically around the resonance mass. The contour lines with a significance  of $5\sigma$ in Figure \ref{contourLim} (a) exhibit a different distribution compared to other  contour lines. This is due to the interference cancellation phenomenon observed in Figure \ref{CrossMuMuToqq} (c) and (e). As the mass of $Z^\prime$ increases while keeping the collision energy a constant, the interference cancellation occurs within the range where $M_{Z^\prime}$ is greater than the collision energy (corresponding to the region where the 
C.M.S energy is less than the resonance peak in Figure~\ref{CrossMuMuToqq}). 
This leads to an increase in the absolute value of contributions from the interference terms involving $Z^\prime$ with photons and $Z^\prime$ with $Z$, as the mixing angle $\sin \alpha$ increases, resulting in a reduction in the cross section. 
This effect causes the contour lines of $5\sigma$ in the region of $\sqrt{s}< M_{Z^\prime}$ differing from the other lines. The cross section can be achieved 0.013 pb at a collision energy of 6000 GeV. 
The region that cannot be detected is limited to the upper right colored region with a significance level less than 5. Figure~\ref{contourLim} (b) illustrates  the sensitivity  of $Z^\prime$ with its mass lower than or equal to the collision energy of 10 TeV. The distribution is similar to the region of $m_{Z^\prime}\le 6000$ GeV in  Figure~\ref{contourLim} (a).   This implies that a muon collider with a  collision energy at 10 TeV  can explore a wide region of $Z^\prime$ mass from 3500 GeV to 10000 GeV with a significance larger than $5\sigma$.
\begin{figure}[tbh] 
\center
	\includegraphics[width=8cm,height=6cm]{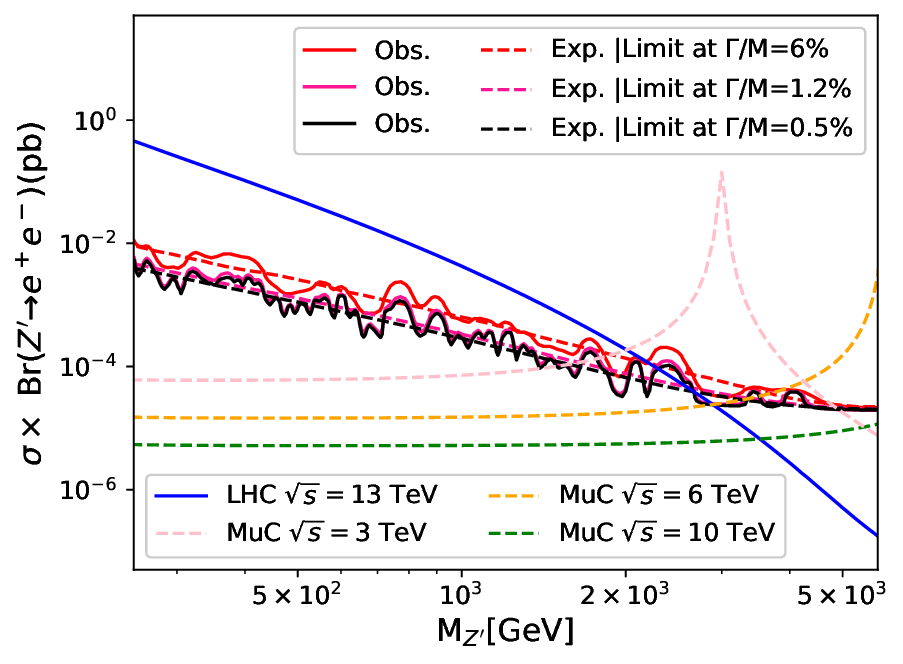}
	\caption{Upper limits on the cross section for $Z^\prime$ production times the branching fraction for $Z^\prime \to e^+ e^-$. The red, deep pink, and black solid lines represent observed values by ATLAS for $\Gamma/M_{Z^\prime}$ equal to 6\%, 1.2\%, and 0.5\%, respectively, with corresponding colored dashed lines representing the expected values for the same parameters~\cite{ATLAS:2019erb,ParticleDataGroup:2022pth}. The blue solid line represents the theoretical value of our model used at the LHC. The pink, orange, and green dashed lines represent theoretical results of our used model with $\eta=0.3$ for the same process at a 3 TeV, 6 TeV, and 10 TeV muon collider, respectively. }
    \label{eeLim}
\end{figure}
The ATLAS collaboration has performed the observation of $Z^\prime$ at the LHC with $Z^\prime \to e^+ e^-$ process. We give the results of $Z^\prime$ boson in HAHM at the LHC with $\sqrt{s}=13~\text{TeV}$ and $\eta=0.3$.  The simulation results are presented by the blue solid line in Figure \ref{eeLim}.  The mass range lower than 2000 GeV is excluded by the LHC experimental results.   
For the muon collider, we have displayed the production of $\mu^+\mu^- \to  e^+ e^-$  at the collision energy of 3, 6, 10 TeV with $\eta=0.3$.  We can find that the excluded mass region can be enlarged with the resonance peak effects. On the other hand, with the collision energy increasing, the study of heavy $Z^\prime$ boson undetected at the LHC can be conducted at the  muon collider.
\section{Summary}
In this paper, we investigate the new neutral gauge boson $Z^\prime$ production and decay at the muon colliders. The interactions and couplings  of $Z^\prime$ with SM fermions and gauge bosons derive from the Hidden Abelian Higgs Model proposed by J. D. Wells.  The free parameters in the model are chosen as the mass of $Z^\prime$ and $\sin\alpha$, the typical value is set as $m_{Z^\prime}=6$ TeV following the collision energy of future muon collider and $\sin\alpha\le 1\times 10^{-4}$ which is safe to guarantee the current experimental measurement of $Z$ boson mass.  The cross section , final state angular distribution and the forward-backward asymmetry are studied in the mainly decay modes of $Z^\prime\rightarrow q\bar{q}, ~t\bar{t},~l^+ l^-,~ ZH, ~W^+W^- $. Each channel has special phenomena for studying properties of $Z^\prime$ boson. The processes of  light quark pair and lepton pair final state have the largest branching fraction for the search of  $Z^\prime$ boson.  With $m_{Z^\prime}=6000$ GeV and  $\sin\alpha \approx 1\times 10^{-4}$, the cross section of  $\mu^+ \mu^-\rightarrow q\bar{q}$ can be reached 12 $\text{pb}$ at the C.M.S. energy of 6 TeV.  The cross section distributions with collision energies show discrepancies around the resonance peak with $Z^\prime$ coupling to up-type or down-type quarks. While the distributions of $\mu^+ \mu^-\rightarrow e^+e^-$ and $\mu^+ \mu^-\rightarrow \mu^+\mu^-$ are different around the resonance peak because of the contribution from different Feynman diagrams.  The  interactions of $Z^\prime$ with gauge boson and Higgs boson can be studied in $\mu^+ \mu^-\rightarrow ZH$ and $\mu^+ \mu^-\rightarrow W^+W^-$ process. Although the branching ratio is a few percent, it shows the potential to test the new gauge interaction at the future muon colliders.  

The angular distribution of the final leptons decaying from $Z^\prime$ boson effectively reveals the  properties of $Z^\prime$ coupling with SM particles. 
The contribution from $Z^\prime$ is obvious in the angular distribution comparing with the SM process. Meanwhile, the angular distributions are also sensitive to the  mixing angle between $Z-Z^\prime$. The corresponding forward-backward asymmetry defined by the angular distribution shows the numerically discrepancy with various parameters in the new physics model. This exotic neutral gauge boson can be introduced in many new physics models. The studies on  $Z^\prime$ boson depend on its interaction with the SM particles. Roughly there are two kinds of sources for the new interaction, one is coming from the SM-like gauge interaction $g^\prime J^{\mu}Z_{\mu}$, and the other is from the kinetic mixing of $Z-Z^\prime$.  The study of this work focuses on the latter one. The production of  $Z^\prime$ boson is very similar in various models except for the production rate proportional to different couplings. The decay of $Z^\prime$ boson is also very general except the specific forbidden models. The mass constraints,  production rates and  detailed distributions will be differing from models depending on the input of the free parameters, but the methods of studying $Z^\prime$ boson in this paper can be adopted to the other models. The studies of $Z^\prime$ boson at the muon collider not only provide the new methods for the search of new physics, but also  reveal the prospect on the new physics model at the future muon collider.  
\section*{Ackonwledgement}
This work was supported by the Natural Science Foundation of Shandong Province under grant No.~ZR2022MA065 and National Natural Science Foundation of China (NNSFC) under grant Nos.12235008.

\bibliographystyle{apsrev4-2}
\bibliography{Reference}
\clearpage
\begin{appendices}
\section*{Appendix A: Detailed study of process $\mu^+ \mu^- \rightarrow q \bar{q}$}\label{appeA}
The vertex  of $Z$ boson in the new physics model can be written as,
\begin{equation}   
\begin{split}
    \bar{\psi} \psi Z &:\frac{ig}{\cos\theta_W}[\cos\alpha(1-\eta \sin\theta_W \tan\alpha)]\gamma^\mu \left[T^3_L - \frac{(1-\eta \tan \alpha/ \sin \theta_W)}{(1-\eta \sin\theta_W \tan\alpha)}\sin^2\theta_W Q \right] \\
    &=\frac{ig}{2\cos \theta_W} \gamma^\mu \left[(\cos\alpha - \eta \cos\alpha \sin\theta_W \tan\alpha)(1-\gamma_5)T_L^3  \right.\\
    &\left.~~~~~~~~~~~~~~~~~~~~~~~~~~~~~~~~~~~~-(\cos\alpha -\eta \cos\alpha \sin\theta_W \tan\alpha)\frac{(1-\eta \tan\alpha/\sin\theta_W)}{(1-\eta \sin\theta_W \tan\alpha)}  \sin^2 \theta_WQ  \right],
\end{split}
\end{equation}
If we require $\alpha=0$ to fulfill the SM, then we obtain the following relationships,
\begin{equation}
\begin{split}
\cos \alpha- \eta \cos\alpha \sin\theta_W \tan\alpha = 1 \\
(\cos \alpha- \eta \cos\alpha \sin\theta_W \tan\alpha)\frac{1- \eta \tan\alpha/\sin\theta_W}{(1-\eta \sin\theta_W \tan\alpha)} =1
\end{split}
\end{equation}
So  equation (\ref{eq3}) can be reduced to,
\begin{equation}
\begin{split}
\bar{\psi} \psi Z &: \frac{ig}{\cos\theta_W} \gamma^{\mu} \left[(1-\gamma_5)T_L^3 -\sin^2\theta_W Q \right] \\
&=\frac{ig}{\cos\theta_W}\gamma^{\mu} \times \frac{1}{2}(c_V^f-c_A^f \gamma^5)
\end{split}
\end{equation}
where $c_V^f = T_L^3 - 2\sin^2 \theta_{W}  Q_f$ and $c_A^f = T_L^3$.
The vertex of $Z^\prime$ and fermions is
\begin{equation}
    \bar{\psi} \psi Z^\prime:\frac{-ig}{\cos\theta_W}\left[\cos\alpha(\tan\alpha+\eta \sin\theta_W)\right] \gamma^\mu \left[T^3_L - \frac{(\tan\alpha+\eta/ \sin \theta_W)}{(\tan\alpha+\eta \sin\theta_W \tan\alpha)}\sin^2\theta_W Q \right].
\end{equation}
It can be shown in the new expression of  
\begin{equation}
\begin{split}
\bar{\psi} \psi Z^{\prime} &: -\frac{ig}{\cos\theta_W}\gamma^\mu \left[(1-\gamma_5)C_1 T^3_L - C_1 C_2\sin^2 \theta_W Q\right] \\
& = -\frac{ig}{\cos\theta_W}\gamma^{\mu} \times \frac{1}{2}(c_V^{f\prime}-c_A^{f\prime} \gamma^5)
\end{split}
\end{equation}
where $c_V^{f\prime} = C_1T_L^3-2C_1 C_2\sin^2\theta_W Q$ and $c_A^{f\prime} = C_2T^3_L$ for
\begin{equation}
\begin{split}
C_1 = \frac{\tan \alpha +\eta/\sin\theta_W}{\tan\alpha+\eta \sin \theta_W} 
,~~C_2 = \cos\alpha \tan\alpha + \eta \cos\alpha \sin \theta_W.
\end{split}
\end{equation}
we list some specific values of $c_V^f$ and $c_A^f$ in SM and new physics Model with different $\eta$ respectively in Table \ref{tb1}. \par

\begin{table}[H]
\begin{center}
\setlength{\abovecaptionskip}{4pt}
\setlength{\belowcaptionskip}{0pt}
\begin{tabular}{ c c c c c c c c c }
\toprule
$f$ & $Q_f$ & $c_A^f(c_A^{f\prime})$ & $c_V^f(c_V^{f\prime})$ & $C_1$ & $C_2$ & $\sin\alpha$ & $\eta$\\ 
\midrule
$\nu_e, \nu_{\mu},...$ & $0$& $0.5$  & $0.5$ & \multirow{4}*{$1$} & \multirow{4}*{$1$} & \multirow{4}*{$0$} & \multirow{4}*{SM}\\  
$e^-,\mu^-,...$ & $-1$ & $-0.5$ & $-0.03$ & ~ & ~ & ~ & ~\\
$u,c,...$ & $0.67$ & $0.5$ & $0.19$ & ~ & ~ & ~ & ~\\
$d,s,...$ & $-0.34$ & $-0.5$ & $-0.34$ & ~ & ~ & ~ & ~\\
\hline
$\nu_e, \nu_{\mu},...$ & $0$& $0.022$  & $0.022$ & \multirow{4}*{$0.043$} & \multirow{4}*{$4.499$}  & \multirow{4}*{$1 \times 10^{-5}$} & \multirow{4}*{0.092}\\  
$e^-,\mu^-,...$ & $-1$ & $-0.022$ & $0.065$ & ~ & ~ & ~ & ~\\
$u,c,...$ & $0.67$ & $0.022$ & $-0.036$ & ~ & ~ & ~ & ~\\
$d,s,...$ & $-0.34$ & $-0.022$ & $0.007$ & ~ & ~ & ~ & ~\\
\cline{1-8}
$\nu_e, \nu_{\mu},...$ & $0$& $0.108$  & $0.108$ & \multirow{4}*{$0.216$} & \multirow{4}*{$4.500$} & \multirow{4}*{$5 \times 10^{-5}$} & \multirow{4}*{0.459}\\  
$e^-,\mu^-,...$ & $-1$ & $-0.108$ & $0.325$ & ~ & ~ & ~ & ~\\
$u,c,...$ & $0.67$ & $0.108$ & $-0.180$ & ~ & ~ & ~ & ~\\
$d,s,...$ & $-0.34$ & $-0.108$ & $0.036$ & ~ & ~ & ~ & ~\\
\cline{1-8}
$\nu_e, \nu_{\mu},...$ & $0$& $0.216$  & $0.216$ & \multirow{4}*{$0.433$} & \multirow{4}*{$4.500$} & \multirow{4}*{$1 \times 10^{-4}$} & \multirow{4}*{0.918}\\  
$e^-,\mu^-,...$ & $-1$ & $-0.216$ & $0.649$ & ~ & ~ & ~ & ~\\
$u,c,...$ & $0.67$ & $0.216$ & $-0.361$ & ~ & ~ & ~ & ~\\
$d,s,...$ & $-0.34$ & $-0.216$ & $0.072$ & ~ & ~ & ~ & ~\\
\bottomrule

\end{tabular}
\caption{Vertex Factors}
\label{tb1}
\end{center}
\end{table}
The total amplitude of process $\mu^+ \mu^- \rightarrow \gamma^*/Z^{\prime} \rightarrow f \bar{f}$ can be written as
\begin{equation}
\mathcal{M}_{total}^2=\mathcal{M}_1^2+\mathcal{M}_2^2+2Re(\mathcal{M}_1 \mathcal{M}_2)
\end{equation}
where $\mathcal{M}_1$ and $\mathcal{M}_2$ are the amplitude produced by $\gamma^*$ and $Z^{\prime}$ as a propagator respectively. In Figure~\ref{CrossMuMuToqq} (e), we can clearly find that when the C.M.S. energy is close to the mass of $Z^\prime$, there is a negative term in the total amplitude. The term with negative value is $2Re(\mathcal{M}_1 \mathcal{M}_2)$  when the C.M.S. energy is close to the mass of $Z^\prime$ and less than the mass of $Z^\prime$. So we can give
\begin{equation}
2Re(\mathcal{M}_1 \mathcal{M}_2) = Re \left(\frac{48e^4Q_q}{s[(s-M^2_{Z^\prime})+i\Gamma_{Z^\prime}M_{Z^\prime}]} \right)\frac{1}{\sin^2 2\theta_W} \left[c^{\mu \prime}_V c^{q \prime}_V(t^2+u^2)-c_A^{\mu \prime} c_A^{q\prime}(t^2-u^2)\right],
\end{equation}
where $t^2+u^2=\frac{s^2}{2}(1+\cos^2 \theta)$ and $t^2-u^2=-s^2\cos\theta$ ($\theta$ is the angle between the outgoing and incoming particles). We rewrite 
\begin{equation}
2Re(\mathcal{M}_1 \mathcal{M}_2) = \frac{-24e^4Q_q s(s-M^2_{Z^\prime})}{(s-M^2_{Z^\prime})^2+\Gamma^2_{Z^{\prime}}M^2_{Z^\prime}}\frac{1}{\sin^2 2\theta_W}\left[c^{\mu \prime}_V c^{q \prime}_V(1+\cos^2\theta)+2c_A^{\mu \prime} c_A^{q\prime} \cos\theta \right],
\end{equation}
\begin{equation}
    \mathcal{M}_2^2=\frac{12e^4 s^2}{[(s-M^2_{Z^\prime})^2+\Gamma^2_{Z^\prime} M^2_{Z^\prime}]\sin^4 2 \theta_W}[c_{VV}^{\mu}c_{VV}^{q}(1+\cos^2 \theta)+2c_{AA}^{\mu}c_{AA}^{q}\cos \theta],
\end{equation}
where $s$ is the square of the C.M.S. energy, $\Gamma_{Z^\prime}$ is the decay width of $Z^\prime$ boson, $Q_q$ is the charge of quarks, $\theta$ is the angle between the initial and final particles moving direction, $c_{VV}^f=(c_V^{f^{\prime}})^2+(c_A^{f^{\prime}})^2$ and $c_{AA}^f=2c_A^{f^{\prime}}c_V^{f^{\prime}}$.
In fact, both the interference between $Z^\prime$ and $Z$ boson, and the interference between $Z^\prime$  and photon contribute negatively to the total scattering cross-section. However, the contribution from the interference of $Z^\prime$ boson and $Z$ boson is small and it does not significantly affect the total scattering cross section.The negative contribution from $[\mathcal{M}^2_2+2Re(\mathcal{M}_1 \mathcal{M}_2)]_{d,s,b}$ is larger than that from  $[\mathcal{M}^2_2+2Re(\mathcal{M}_1 \mathcal{M}_2)]_{u,c,t}$, as observed in Figure~\ref{CrossMuMuToqq} (c), while this phenomenon is not prominent in Figure~\ref{CrossMuMuToqq} (d).
\section*{Appendix B: The cross section of process $\mu^+ \mu^- \rightarrow W^+ W^-$}\label{appeB}
 The Feynman rule of weak interaction with the vertices of $WW\gamma^*$, $WWZ$ and $WWZ^\prime$ are given,
\begin{equation} \label{eq17}
\tikzset{every picture/.style={line width=0.75pt},baseline=(current bounding box.center)} 
\begin{tikzpicture}
[x=0.75pt,y=0.75pt,yscale=-1,xscale=1]
\draw    (136.67,73.47) .. controls (134.5,74.37) and (132.96,73.73) .. (132.05,71.56) .. controls (131.15,69.38) and (129.61,68.74) .. (127.43,69.65) .. controls (125.26,70.56) and (123.72,69.92) .. (122.81,67.75) .. controls (121.9,65.57) and (120.36,64.93) .. (118.18,65.84) .. controls (116,66.75) and (114.46,66.11) .. (113.56,63.93) .. controls (112.65,61.76) and (111.11,61.12) .. (108.94,62.03) .. controls (106.76,62.94) and (105.22,62.3) .. (104.32,60.12) .. controls (103.42,57.94) and (101.88,57.3) .. (99.7,58.21) .. controls (97.53,59.12) and (95.98,58.48) .. (95.07,56.31) .. controls (94.17,54.13) and (92.63,53.49) .. (90.45,54.4) .. controls (88.27,55.31) and (86.73,54.67) .. (85.83,52.49) .. controls (84.92,50.32) and (83.38,49.68) .. (81.21,50.59) -- (77.2,48.94) -- (77.2,48.94) ;
\draw    (197,48.94) .. controls (196.09,51.11) and (194.54,51.73) .. (192.37,50.82) .. controls (190.2,49.9) and (188.65,50.53) .. (187.74,52.7) .. controls (186.83,54.87) and (185.28,55.5) .. (183.11,54.59) .. controls (180.94,53.67) and (179.39,54.3) .. (178.48,56.47) .. controls (177.57,58.64) and (176.02,59.27) .. (173.85,58.35) .. controls (171.68,57.44) and (170.13,58.07) .. (169.21,60.24) .. controls (168.3,62.41) and (166.75,63.04) .. (164.58,62.12) .. controls (162.41,61.2) and (160.86,61.83) .. (159.95,64) .. controls (159.04,66.17) and (157.49,66.8) .. (155.32,65.89) .. controls (153.15,64.97) and (151.6,65.6) .. (150.69,67.77) .. controls (149.78,69.94) and (148.23,70.57) .. (146.06,69.65) .. controls (143.89,68.74) and (142.34,69.37) .. (141.42,71.54) .. controls (140.51,73.71) and (138.96,74.34) .. (136.79,73.42) -- (136.67,73.47) -- (136.67,73.47) ;
\draw    (136.85,125.91) .. controls (135.18,124.25) and (135.17,122.58) .. (136.83,120.91) .. controls (138.49,119.24) and (138.48,117.57) .. (136.81,115.91) .. controls (135.14,114.24) and (135.14,112.58) .. (136.8,110.91) .. controls (138.46,109.24) and (138.45,107.57) .. (136.78,105.91) .. controls (135.11,104.25) and (135.1,102.58) .. (136.76,100.91) .. controls (138.42,99.24) and (138.42,97.58) .. (136.75,95.91) .. controls (135.08,94.25) and (135.07,92.58) .. (136.73,90.91) .. controls (138.39,89.24) and (138.38,87.57) .. (136.71,85.91) .. controls (135.04,84.24) and (135.04,82.58) .. (136.7,80.91) .. controls (138.36,79.24) and (138.35,77.57) .. (136.68,75.91) -- (136.67,73.47) -- (136.67,73.47) ;
\draw [shift={(136.67,73.47)}, rotate = 269.81] [color={rgb, 255:red, 0; green, 0; blue, 0 }  ][fill={rgb, 255:red, 0; green, 0; blue, 0 }  ][line width=0.75]      (0, 0) circle [x radius= 2.68, y radius= 2.68]   ;
\draw    (127.36,98.63) -- (127.36,120.63) ;
\draw [shift={(127.36,96.63)}, rotate = 90] [fill={rgb, 255:red, 0; green, 0; blue, 0 }  ][line width=0.08]  [draw opacity=0] (12,-3) -- (0,0) -- (12,3) -- cycle    ;
\draw    (183.28,62.62) -- (160.12,72.89) ;
\draw [shift={(185.11,61.81)}, rotate = 156.09] [fill={rgb, 255:red, 0; green, 0; blue, 0 }  ][line width=0.08]  [draw opacity=0] (12,-3) -- (0,0) -- (12,3) -- cycle    ;
\draw    (86.94,60.29) -- (111.85,72.1) ;
\draw [shift={(85.13,59.43)}, rotate = 25.36] [fill={rgb, 255:red, 0; green, 0; blue, 0 }  ][line width=0.08]  [draw opacity=0] (12,-3) -- (0,0) -- (12,3) -- cycle    ;

\draw (129.36,124.03) node [anchor=north west][inner sep=0.75pt]  [font=\scriptsize]  {$\gamma ^{*}$};
\draw (195.66,32.31) node [anchor=north west][inner sep=0.75pt]  [font=\scriptsize]  {$W^{-}$};
\draw (60.27,32.31) node [anchor=north west][inner sep=0.75pt]  [font=\scriptsize]  {$W^{+}$};
\draw (85.17,40.06) node [anchor=north west][inner sep=0.75pt]  [font=\scriptsize]  {$\nu $};
\draw (175.66,40.85) node [anchor=north west][inner sep=0.75pt]  [font=\scriptsize]  {$\mu $};
 \draw (139.46,115.24) node [anchor=north west][inner sep=0.75pt]  [font=\scriptsize]  {$\lambda $};
\draw (111.88,105.75) node [anchor=north west][inner sep=0.75pt]  [font=\scriptsize]  {$q$};
\draw (87.48,68.72) node [anchor=north west][inner sep=0.75pt]  [font=\scriptsize]  {$k_{-}$};
\draw (166.7,69.51) node [anchor=north west][inner sep=0.75pt]  [font=\scriptsize]  {$k_{+}$};
\draw (211,69.4) node [anchor=north west][inner sep=0.75pt]    {$=ie\left[ g^{\mu \nu }( k_{-} -k_{+})^{\lambda } +g^{\nu \lambda }( -q-k_{-})^{\mu } +g^{\lambda \mu }( q+k_{+})^{\nu }\right]$};

\end{tikzpicture},
\end{equation}
\begin{equation}\label{eq18}
\tikzset{every picture/.style={line width=0.75pt},baseline=(current bounding box.center)} 
\begin{tikzpicture}[x=0.75pt,y=0.75pt,yscale=-1,xscale=1]

\draw    (136.67,73.47) .. controls (134.5,74.37) and (132.96,73.73) .. (132.05,71.56) .. controls (131.15,69.38) and (129.61,68.74) .. (127.43,69.65) .. controls (125.26,70.56) and (123.72,69.92) .. (122.81,67.75) .. controls (121.9,65.57) and (120.36,64.93) .. (118.18,65.84) .. controls (116,66.75) and (114.46,66.11) .. (113.56,63.93) .. controls (112.65,61.76) and (111.11,61.12) .. (108.94,62.03) .. controls (106.76,62.94) and (105.22,62.3) .. (104.32,60.12) .. controls (103.42,57.94) and (101.88,57.3) .. (99.7,58.21) .. controls (97.53,59.12) and (95.98,58.48) .. (95.07,56.31) .. controls (94.17,54.13) and (92.63,53.49) .. (90.45,54.4) .. controls (88.27,55.31) and (86.73,54.67) .. (85.83,52.49) .. controls (84.92,50.32) and (83.38,49.68) .. (81.21,50.59) -- (77.2,48.94) -- (77.2,48.94) ;
\draw    (197,48.94) .. controls (196.09,51.11) and (194.54,51.73) .. (192.37,50.82) .. controls (190.2,49.9) and (188.65,50.53) .. (187.74,52.7) .. controls (186.83,54.87) and (185.28,55.5) .. (183.11,54.59) .. controls (180.94,53.67) and (179.39,54.3) .. (178.48,56.47) .. controls (177.57,58.64) and (176.02,59.27) .. (173.85,58.35) .. controls (171.68,57.44) and (170.13,58.07) .. (169.21,60.24) .. controls (168.3,62.41) and (166.75,63.04) .. (164.58,62.12) .. controls (162.41,61.2) and (160.86,61.83) .. (159.95,64) .. controls (159.04,66.17) and (157.49,66.8) .. (155.32,65.89) .. controls (153.15,64.97) and (151.6,65.6) .. (150.69,67.77) .. controls (149.78,69.94) and (148.23,70.57) .. (146.06,69.65) .. controls (143.89,68.74) and (142.34,69.37) .. (141.42,71.54) .. controls (140.51,73.71) and (138.96,74.34) .. (136.79,73.42) -- (136.67,73.47) -- (136.67,73.47) ;
\draw    (136.85,125.91) .. controls (135.18,124.25) and (135.17,122.58) .. (136.83,120.91) .. controls (138.49,119.24) and (138.48,117.57) .. (136.81,115.91) .. controls (135.14,114.24) and (135.14,112.58) .. (136.8,110.91) .. controls (138.46,109.24) and (138.45,107.57) .. (136.78,105.91) .. controls (135.11,104.25) and (135.1,102.58) .. (136.76,100.91) .. controls (138.42,99.24) and (138.42,97.58) .. (136.75,95.91) .. controls (135.08,94.25) and (135.07,92.58) .. (136.73,90.91) .. controls (138.39,89.24) and (138.38,87.57) .. (136.71,85.91) .. controls (135.04,84.24) and (135.04,82.58) .. (136.7,80.91) .. controls (138.36,79.24) and (138.35,77.57) .. (136.68,75.91) -- (136.67,73.47) -- (136.67,73.47) ;
\draw [shift={(136.67,73.47)}, rotate = 269.81] [color={rgb, 255:red, 0; green, 0; blue, 0 }  ][fill={rgb, 255:red, 0; green, 0; blue, 0 }  ][line width=0.75]      (0, 0) circle [x radius= 2.68, y radius= 2.68]   ;
\draw    (127.36,98.63) -- (127.36,120.63) ;
\draw [shift={(127.36,96.63)}, rotate = 90] [fill={rgb, 255:red, 0; green, 0; blue, 0 }  ][line width=0.08]  [draw opacity=0] (12,-3) -- (0,0) -- (12,3) -- cycle    ;
\draw    (183.28,62.62) -- (160.12,72.89) ;
\draw [shift={(185.11,61.81)}, rotate = 156.09] [fill={rgb, 255:red, 0; green, 0; blue, 0 }  ][line width=0.08]  [draw opacity=0] (12,-3) -- (0,0) -- (12,3) -- cycle    ;
\draw    (86.94,60.29) -- (111.85,72.1) ;
\draw [shift={(85.13,59.43)}, rotate = 25.36] [fill={rgb, 255:red, 0; green, 0; blue, 0 }  ][line width=0.08]  [draw opacity=0] (12,-3) -- (0,0) -- (12,3) -- cycle    ;

\draw (128.36,129.03) node [anchor=north west][inner sep=0.75pt]  [font=\scriptsize]  {$Z$};
\draw (195.66,32.31) node [anchor=north west][inner sep=0.75pt]  [font=\scriptsize]  {$W^{-}$};
\draw (60.27,32.31) node [anchor=north west][inner sep=0.75pt]  [font=\scriptsize]  {$W^{+}$};
\draw (85.17,40.06) node [anchor=north west][inner sep=0.75pt]  [font=\scriptsize]  {$\nu $};
\draw (175.66,40.85) node [anchor=north west][inner sep=0.75pt]  [font=\scriptsize]  {$\mu $};
\draw (139.46,115.24) node [anchor=north west][inner sep=0.75pt]  [font=\scriptsize]  {$\lambda $};
\draw (111.88,105.75) node [anchor=north west][inner sep=0.75pt]  [font=\scriptsize]  {$q$};
\draw (87.48,68.72) node [anchor=north west][inner sep=0.75pt]  [font=\scriptsize]  {$k_{-}$};
\draw (166.7,69.51) node [anchor=north west][inner sep=0.75pt]  [font=\scriptsize]  {$k_{+}$};
\draw (211,69.4) node [anchor=north west][inner sep=0.75pt]    {$=ig\cos \theta _{w}\cos \alpha \left[ g^{\mu \nu }( k_{-} -k_{+})^{\lambda } +g^{\nu \lambda }( -q-k_{-})^{\mu } +g^{\lambda \mu }( q+k_{+})^{\nu }\right]$};
\end{tikzpicture},
\end{equation}

\begin{equation}\label{eq19}
\tikzset{every picture/.style={line width=0.75pt},baseline=(current bounding box.center)} 
\begin{tikzpicture}[x=0.75pt,y=0.75pt,yscale=-1,xscale=1]

\draw    (136.67,73.47) .. controls (134.5,74.37) and (132.96,73.73) .. (132.05,71.56) .. controls (131.15,69.38) and (129.61,68.74) .. (127.43,69.65) .. controls (125.26,70.56) and (123.72,69.92) .. (122.81,67.75) .. controls (121.9,65.57) and (120.36,64.93) .. (118.18,65.84) .. controls (116,66.75) and (114.46,66.11) .. (113.56,63.93) .. controls (112.65,61.76) and (111.11,61.12) .. (108.94,62.03) .. controls (106.76,62.94) and (105.22,62.3) .. (104.32,60.12) .. controls (103.42,57.94) and (101.88,57.3) .. (99.7,58.21) .. controls (97.53,59.12) and (95.98,58.48) .. (95.07,56.31) .. controls (94.17,54.13) and (92.63,53.49) .. (90.45,54.4) .. controls (88.27,55.31) and (86.73,54.67) .. (85.83,52.49) .. controls (84.92,50.32) and (83.38,49.68) .. (81.21,50.59) -- (77.2,48.94) -- (77.2,48.94) ;
\draw    (197,48.94) .. controls (196.09,51.11) and (194.54,51.73) .. (192.37,50.82) .. controls (190.2,49.9) and (188.65,50.53) .. (187.74,52.7) .. controls (186.83,54.87) and (185.28,55.5) .. (183.11,54.59) .. controls (180.94,53.67) and (179.39,54.3) .. (178.48,56.47) .. controls (177.57,58.64) and (176.02,59.27) .. (173.85,58.35) .. controls (171.68,57.44) and (170.13,58.07) .. (169.21,60.24) .. controls (168.3,62.41) and (166.75,63.04) .. (164.58,62.12) .. controls (162.41,61.2) and (160.86,61.83) .. (159.95,64) .. controls (159.04,66.17) and (157.49,66.8) .. (155.32,65.89) .. controls (153.15,64.97) and (151.6,65.6) .. (150.69,67.77) .. controls (149.78,69.94) and (148.23,70.57) .. (146.06,69.65) .. controls (143.89,68.74) and (142.34,69.37) .. (141.42,71.54) .. controls (140.51,73.71) and (138.96,74.34) .. (136.79,73.42) -- (136.67,73.47) -- (136.67,73.47) ;
\draw    (136.85,125.91) .. controls (135.18,124.25) and (135.17,122.58) .. (136.83,120.91) .. controls (138.49,119.24) and (138.48,117.57) .. (136.81,115.91) .. controls (135.14,114.24) and (135.14,112.58) .. (136.8,110.91) .. controls (138.46,109.24) and (138.45,107.57) .. (136.78,105.91) .. controls (135.11,104.25) and (135.1,102.58) .. (136.76,100.91) .. controls (138.42,99.24) and (138.42,97.58) .. (136.75,95.91) .. controls (135.08,94.25) and (135.07,92.58) .. (136.73,90.91) .. controls (138.39,89.24) and (138.38,87.57) .. (136.71,85.91) .. controls (135.04,84.24) and (135.04,82.58) .. (136.7,80.91) .. controls (138.36,79.24) and (138.35,77.57) .. (136.68,75.91) -- (136.67,73.47) -- (136.67,73.47) ;
\draw [shift={(136.67,73.47)}, rotate = 269.81] [color={rgb, 255:red, 0; green, 0; blue, 0 }  ][fill={rgb, 255:red, 0; green, 0; blue, 0 }  ][line width=0.75]      (0, 0) circle [x radius= 2.68, y radius= 2.68]   ;
\draw    (127.36,98.63) -- (127.36,120.63) ;
\draw [shift={(127.36,96.63)}, rotate = 90] [fill={rgb, 255:red, 0; green, 0; blue, 0 }  ][line width=0.08]  [draw opacity=0] (12,-3) -- (0,0) -- (12,3) -- cycle    ;
\draw    (183.28,62.62) -- (160.12,72.89) ;
\draw [shift={(185.11,61.81)}, rotate = 156.09] [fill={rgb, 255:red, 0; green, 0; blue, 0 }  ][line width=0.08]  [draw opacity=0] (12,-3) -- (0,0) -- (12,3) -- cycle    ;
\draw    (86.94,60.29) -- (111.85,72.1) ;
\draw [shift={(85.13,59.43)}, rotate = 25.36] [fill={rgb, 255:red, 0; green, 0; blue, 0 }  ][line width=0.08]  [draw opacity=0] (12,-3) -- (0,0) -- (12,3) -- cycle    ;

\draw (130.36,129.03) node [anchor=north west][inner sep=0.75pt]  [font=\scriptsize]  {$Z^{\prime }$};
\draw (195.66,32.31) node [anchor=north west][inner sep=0.75pt]  [font=\scriptsize]  {$W^{-}$};
\draw (60.27,32.31) node [anchor=north west][inner sep=0.75pt]  [font=\scriptsize]  {$W^{+}$};
\draw (85.17,40.06) node [anchor=north west][inner sep=0.75pt]  [font=\scriptsize]  {$\nu $};
\draw (175.66,40.85) node [anchor=north west][inner sep=0.75pt]  [font=\scriptsize]  {$\mu $};
\draw (139.46,115.24) node [anchor=north west][inner sep=0.75pt]  [font=\scriptsize]  {$\lambda $};
\draw (111.88,105.75) node [anchor=north west][inner sep=0.75pt]  [font=\scriptsize]  {$q$};
\draw (87.48,68.72) node [anchor=north west][inner sep=0.75pt]  [font=\scriptsize]  {$k_{-}$};
\draw (166.7,69.51) node [anchor=north west][inner sep=0.75pt]  [font=\scriptsize]  {$k_{+}$};
\draw (211,69.4) node [anchor=north west][inner sep=0.75pt]    {$=-ig\cos \theta _{w}\sin \alpha \left[ g^{\mu \nu }( k_{-} -k_{+})^{\lambda } +g^{\nu \lambda }( -q-k_{-})^{\mu } +g^{\lambda \mu }( q+k_{+})^{\nu }\right]$};
\end{tikzpicture}.
\end{equation}
We assume that $\mathcal{R}$ is the coupling relative to the corresponding the SM. We can find that $\mathcal{R}_{\gamma^* W^+ W^-}=1$, $\mathcal{R}_{Z W^+ W^-}=\cos\alpha$ and $\mathcal{R}_{Z^\prime W^+ W^-}=-\sin\alpha$ by Equations~\eqref{eq17}, \eqref{eq18} and \eqref{eq19}. But we will normally assume rather small kinetic mixing and  $\cos \alpha \approx 1$ and $\sin \alpha \ll 1$. We have calculated the scattering cross section expression of this process through FeynArts and FeynClac. The cross section of $\mu^+ \mu^- \rightarrow W^+ W^-$ can be expressed as
\begin{align*}\label{eq35}
\sigma(&\mu^+ \mu^- \rightarrow W^+ W^-)=\left[\frac{\pi \alpha^2 \text{csc}^4\theta_W}{96s^3 M_W^4(M_W^2-s\cdot\cos^2\theta_W)^2(s-M^2_{Z^\prime})^2}\right]\cdot \Bigg\{s\beta \bigg\{-384(M^2_{Z^\prime} \\
&-s\cos^2\alpha)^2M_W^{10}-8s\big\{2(33\cos2\alpha+7)M_{Z^\prime}^4+s[8(9\cos2\alpha+26)\sin^2\alpha+5(3\eta^2\sin^2 2\alpha\\
&-32)]M_{Z^\prime}^2+2s^2 \cos^2\alpha(17\cos2\alpha+23)\big\}M_W^8+2s^2\big\{4[(51\cos2\alpha+163)\sin^2\alpha\\
&+15\eta^2 \sin^2 2\alpha-4]M^4_{Z^\prime}+s(816\sin^4 \alpha-808\sin^2\alpha-85\eta^2\sin^2 2\alpha+32)M^2_{Z^\prime}\\
&+16s^2\cos^2\alpha(2\cos2\alpha-3)\big\}M_W^6+s^3[2(9578\sin^4\alpha+4\sin^2 \alpha+85\eta^2\sin^2 2\alpha-30)M^4_{Z^\prime}\\
&+8s(-48\sin^4\alpha+5\eta^2\sin^2 2\alpha+15)M^2_{Z^\prime}+s^2(\cos4\alpha-61)]M_W^4+s^3\sin^2 \alpha M^2_{Z^\prime}\Big\{s[5\eta^2\\
&+(5\eta^2+12)\cos2\alpha-8]-4[20\eta^2+(20\eta^2+34)\cos2\alpha-33]M^2_{Z^\prime}\Big\}M_W^2+\frac{1}{2}s^5(34\sin^4\alpha \\
&-5\eta^2\sin^2 2\alpha)M^4_{Z^\prime}\bigg\}\sin^4\theta_W-s\beta\bigg\{24 \Big\{-4M^4_{Z^\prime}+2s(3\cos2\alpha+1)M^2_{Z^\prime}+s^2\cos^2\alpha[5\eta^2\\
&-(5\eta^2+4)\cos2\alpha]\Big\}M_W^{10}+s\Big\{-8(21\cos2\alpha+11)M^4_{Z^\prime}+4s[-15\eta^2+48\cos2\alpha+15(\eta^2\\
&+1)\cos4\alpha+65]M^2_{Z^\prime}+s^2[17(5\eta^2+4)\sin^2 2\alpha-256\cos^2\alpha]\Big\}M^8_W+s^2\Big\{-2[-15\eta^2\\
&+86\cos2\alpha+3(5\eta^2+6)\cos4\alpha-80]M^4_{Z^\prime}+s[-85\eta^2+28\cos2\alpha+85(\eta^2+1)\cos4\alpha\\
&-17]M^2_{Z^\prime}+2s^2[-5\eta^2-42\cos2\alpha+(5\eta^2+4)\cos4\alpha+14]\Big\} M_W^6+s^3\Big\{[4(51\cos2\alpha \tag{26}\\
&-14)\sin^2\alpha+85\eta^2\sin^2 2\alpha-128]M^4_{Z^\prime}+4s[39\cos2\alpha+5(\eta^2-\eta^2\cos4\alpha-\cos4\alpha+6)]\\
&\cdot M^2_{Z^\prime}+s^2[\sin^4\alpha+(4-5\eta^2\cos^2\alpha)\sin^2\alpha-128]\Big\}M^4_W+s^4\sin^2\alpha M^2_{Z^\prime}\Big\{s[5\eta^2+5(\eta^2\\
\displaybreak
&+1)\cos2\alpha-11]-2[20\eta^2+4(5\eta^2+6)\cos2\alpha-27]M^2_{Z^\prime}\Big\}M^2_W+s^5\sin^2\alpha(6\sin^2\alpha\\
&-5\eta^2\cos^2\alpha)M^4_{Z^\prime}\bigg\}\sin^2\theta_W+s^2\beta(12M_W^4+20sM_W^2+s^2)\bigg\{8s^2\beta^2\sin^4\alpha M^4_{Z^\prime}\sin^7\theta_W\\
&+12s^2\beta^2\eta\cos\alpha \sin^3 \alpha M^4_{Z^\prime}\sin^6 \theta_W+\sin^2\alpha M^2_{Z^\prime}\Big\{64(s\cos^2\alpha-M^2_{Z^\prime})M^4_W-2s[(5\eta^2\\
&+5\eta^2\cos2\alpha+20\cos2\alpha-24)M^2_{Z^\prime}+4s\cos2\alpha]M^2_W+5s^2[\eta^2-(\eta^2+4)\sin^2\alpha]M^2_{Z^\prime}\Big\}\\
&\cdot\sin^5\theta_W+2\zeta M^2_{Z^\prime}\Big\{24(s\cos2\alpha-M^2_{Z^\prime})M^4_W+2s[(15-13\cos2\alpha)M^2_{Z^\prime}+s-3s\cos2\alpha]M^2_W\\
&-13s^2\sin^2\alpha M^2_{Z^\prime}\Big\}\sin^4\theta_W-2\zeta\Big\{24(M^2_{Z^\prime}-s\cos^2\alpha)M_W^6+[-26M^4_{Z^\prime}+2s(14\cos2\alpha\\
&-3)M^2_{Z^\prime}+s^2(3\cos2\alpha+1)]M^4_W+sM^2_{Z^\prime}[(19-16\cos2\alpha)M^2_{Z^\prime}+s(4-7\cos2\alpha)]M^2_W\\
&-8s^2\sin^2\alpha M^4_{Z^\prime}\Big\}\sin^2\theta_W-2\zeta(M^2_W-M^2_{Z^\prime})\Big\{2(s\cos2\alpha-M^2_{Z^\prime})M^4_W+s[(1-2\cos2\alpha)M^2_{Z^\prime}\\
&+s(\sin^2\alpha+1)]M^2_W-s^2\sin^2\alpha M^2_{Z^\prime}\Big\}\bigg\}\sin\theta_W-24\text{log}\left(\frac{-2M^2_W+s-s\beta}{-2M^2_W+s+s\beta}\right)M^4_W(M^2_W\\
&-s\cos^2\theta_W)(s-M^2_{Z^\prime})\bigg\{-2[2\sin^2\theta_W M^2_{Z^\prime}+s(\cos2\theta_W\cos^2\alpha+\zeta\sin\theta_W-1)]M^6_W\\
&+2s\Big\{s(2\cos2\theta_W\sin^2\alpha+5\sin^2\theta_W-2\zeta\sin\theta_W-2)-\big\{\cos^2\theta_W \cos2\theta_W \sin^2\alpha+\sin\theta_W\\
&\cdot[\sin\theta_W(\zeta\sin\theta_W+5)-\zeta]-2\big\}M^2_{Z^\prime}\Big\}M^4_W+s^2\Big\{[4\zeta \sin\theta_W\cos^2\theta_W+\cos2\theta_W(1\\
&-4\cos^2\theta_W\sin^2\alpha)]\Big\}M^2_W+s^3\cos^2\theta_W(M^2_{Z^\prime}-s)\bigg\}+s^2\beta\Big\{48\sin^2\alpha(s\cos^2\alpha-M^2_{Z^\prime})M^{10}_W\\
&+4[-6(\cos2\alpha+3)M^4_{Z^\prime}+s(7\cos2\alpha+3\cos4\alpha+38)M^2_{Z^\prime}+s^2(-17\sin^4\alpha+26\sin^2\alpha\\
&-24)]M^8_W+s[(-28\cos2\alpha-6\cos4\alpha+94)M^4_{Z^\prime}+s(6\cos2\alpha+17\cos4\alpha-143)M^2_{Z^\prime}\\
&+2s^2(-15\cos2\alpha+\cos4\alpha+44)]M^6_W+\frac{1}{8}s^2[(96\cos2\alpha-68\cos4\alpha-532)M^4_{Z^\prime}+8s(39\cos2\alpha\\
&-4\cos4\alpha+91)M^2_{Z^\prime}+s^2(-12\cos2\alpha+\cos4\alpha-493)]M^4_W+s^3\sin^2\alpha M^2_{Z^\prime}[(10-8\cos2\alpha)M^2_{Z^\prime}\\
&+s(\cos2\alpha-3)]M^2_W+s^4\sin^4\alpha M^4_{Z^\prime}\Big\}\Bigg\},
\end{align*}
where $\alpha$ is the mixing angle, $\beta$ is the velocity of the $W$ boson, $\zeta=\eta\sin\alpha \cos\alpha $ and $\beta=(1-4M^2_W/s)^{1/2}$.

\end{appendices}

\end{document}